\documentclass[review,3p,sort&compress]{elsarticle}
\usepackage{amssymb}
\usepackage{siunitx}
\usepackage{color,soul}
\usepackage{graphicx}
\usepackage{amsthm}
\usepackage{hyperref}
\usepackage{amsmath}
\usepackage{float}
\hypersetup{
    colorlinks =    true,
    linkcolor =     [rgb]{0.0,0.0,1.0},
   anchorcolor =   [rgb]{0.0,0.0,1.0},
    citecolor =     [rgb]{0,0,1},
    filecolor =     [rgb]{0.0,0.0,1.0},
    urlcolor =      [rgb]{0.0,0.0,1.0},
}
\usepackage[nameinlink]{cleveref}
\crefname{figure}{Figure}{Figures}
\crefname{equation}{Equation}{Equations}
\crefname{table}{Table}{Tables}
\usepackage{caption}
\usepackage{siunitx}
\usepackage{subcaption}
\usepackage{svg}
\journal{arXiv}
\begin{document}
\begin{frontmatter}
\title{Study of triaxial loading of segregated granular assemblies through experiments and DEM simulations}
\author[inst1]{Venakata Rama Manoj Pola}
\affiliation[inst1]{organization={Department of Mechanical Engineering, Indian Institute of Technology Madras},
city={Chennai},
postcode={600036}, 
state={Tamil Nadu},
country={India}}
\affiliation[inst2]{organization={Center for Soft and Biological Matter, Indian Institute of Technology Madras},
            city={Chennai},
            postcode={600036}, 
            state={Tamil Nadu},
            country={India}}
\cortext[cor1]{Corresponding author}
\author[inst1,inst2]{Ratna Kumar Annabattula\corref{cor1}}
\ead{ratna@iitm.ac.in}
\begin{abstract}
A simple position-dependent body force-based confinement for simulating triaxial tests using the Discrete Element Method is presented. The said method is used to perform triaxial simulations on mono-disperse and segregated assemblies of glass spheres. The macroscopic load response obtained in simulations is validated with experimental load response. A mesh construction algorithm is presented to check whether the confinement applied in the triaxial simulations is accurate. The particle displacement data obtained from triaxial simulations are used to obtain a particle-wise average strain tensor. This is further used to compare the strain localisation between the mono-disperse and segregated assemblies. It is observed that, in the segregated assembly, the interface between the two particle phases acts as a barrier for strain localisation, and the smaller particles preferentially undergo a higher degree of shear strain on average. 
\end{abstract}

\begin{keyword}
granular materials \sep segregated \sep discrete element method \sep triaxial test \sep shear strain \sep flexible membrane
\end{keyword}

\end{frontmatter}

\section{Introduction}
\label{sec:Introduction}
 Discrete Element Method(DEM)~\cite{cundall1979discrete} is a numerical tool to simulate granular materials. Most widely used for DEM treats individual particles as distinct free-moving bodies whose state~(positions and velocities) are updated at regular intervals. To update the state of the particles, the forces on the particles are calculated at the beginning of each interval. The forces on the particles are obtained either through contact with other bodies or through a force field. Various contact force models have been used to predict the force transmitted through contact between two particles or between a particle and geometry~\cite{Hertz1882, Mindlin1949, Mindlin1953, SAKAGUCHI1993}. After the force is obtained, the net acceleration vector of the particle is calculated using Newton's second law of motion. This acceleration vector is temporally integrated to obtain the updated velocities and positions of the particles at the subsequent time step. This process is continued for the desired amount of time. 
 
 Triaxial tests have been simulated in DEM using many different approaches. The main challenge in simulating triaxial tests is accurately modelling the membrane and confinement pressure. The most common approach is to model the membrane with bonded particles and construct area elements with the membrane particles as nodes to calculate the force required for confinement~\cite{Qin2021,Fazekas2006,deBono2012,Wu2021,Wang2008}. Another approach is to model the membrane with servo-controlled geometric elements to maintain the confinement~\cite{Lee2012}. A variation of this method is to use a rigid cylinder whose radius can be controlled to maintain confinement~\cite{Wu2017,Wu2017_2}. More advanced approaches include modelling the membrane with a finite element mesh and coupling DEM with FEA to simulate confinement~\cite{Ma2014} and modelling the membrane with deformable discrete elements called PFacets~\cite{Yang2019}. 
 
 In this paper, the membrane is modelled with bonded particles, as has been done before. However, confinement is applied as a position-dependent body force on the particles comprising the membrane. There are two main advantages of this approach. Firstly, it is relatively uncomplicated to incorporate into DEM, as additional code is not required to track area elements. Secondly, it is computationally less expensive for the same reason, i.e., additional computational power is not required to track a mesh. There are some drawbacks to this approach, too. Firstly, due to the nature of the confinement force, we can only simulate triaxial tests with a non-rotatable platen. This is because the inherent assumption in the approach is that the membrane particles move in an axis-symmetric manner. Such axis-symmetric deformation is only seen when the platens used for loading are non-rotatable. Also, large lateral deformations induce erroneous extra confinement onto the specimen. This is because of another approximation: the configuration of the membrane is such that any area-element vector on the membrane will be radially outwards from the membrane. This approximation can only be made when the membrane has not deformed much.

 Segregation is a prevalent phenomenon in granular materials~\cite{ottino2000mixing}. The sources of the segregation can be many, including flow, vibration, erosion, etc. In this study, the cause of initiation or persistence of the segregation is not examined. Only the effect of segregation on the material's behaviour when subjected to triaxial loads is examined. Experimental studies have shown a correlation between segregation and strength of granular assemblies~\cite{yu2021influence}. In this paper, we focus on the pattern of deformation. We use DEM to simulate triaxial loading of segregated specimens and obtain strain from the DEM data~\cite{thomas1997discontinuous,o2011particulate}. As will be seen, a counter-intuitive pattern emerges as the specimens are loaded. We see that irrespective of the size of the larger particles, the smaller particles preferentially undergo a larger degree of lateral displacement and, as an implication, a higher degree of shear strain. 
\section{Triaxial test}
\label{sec: triaxial experiments}
\begin{figure}[htbp!]
         \centering
         \includegraphics[width=\textwidth]{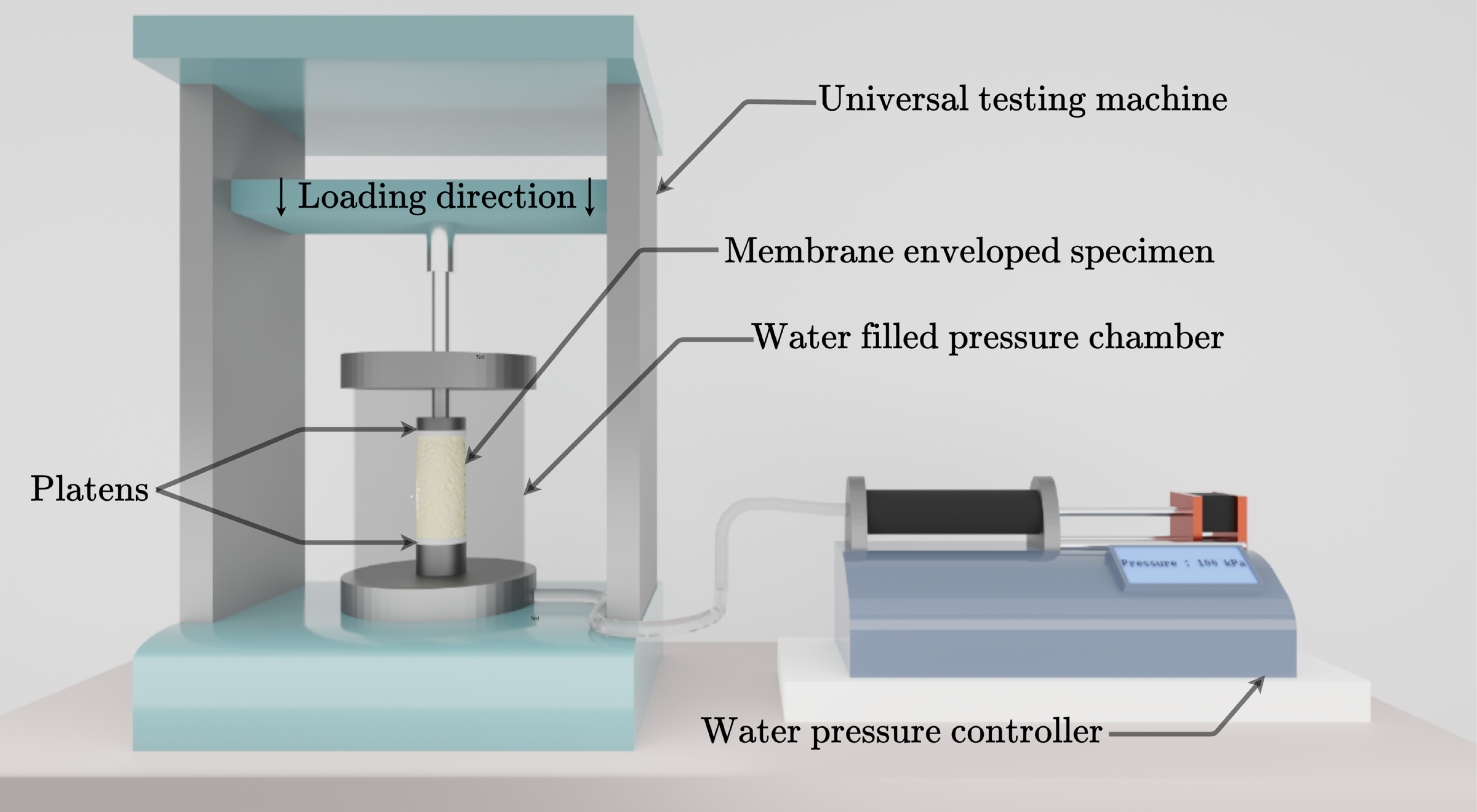}
        \caption{Typical triaxial experimental setup. }
        \label{fig:setup}
\end{figure}

Triaxial tests on granular materials are performed on cylindrical specimens held in a thin latex membrane. The membrane is tightly attached to two platens, one at the bottom and one at the top. The bottom platen is fixed, and the top platen can move in the axial direction. The entire setup is placed in a chamber filled with water and subsequently pressurised. A simplified illustration of a triaxial setup can be seen in \cref{fig:setup}. 
Standard \SI{38}{mm} specimens comprised of soda-lime glass spheres were used for the test. Initial packing density plays a critical role in the force-displacement response and rheology of the granular assembly. In order to maintain uniformity throughout, all the tests were carried out with 70 grams of the glass beads to obtain an average packing density of 60\%. The sample was loaded by moving the top platen down at a rate of \SI[per-mode=symbol]{5}{\milli\meter\per\minute}, which corresponds to a strain rate of \SI{0.003125}{\per\second}. This strain rate was chosen to reduce the compute time in the corresponding simulations. Shimadzu AGS-X\textsuperscript{\texttrademark} series universal testing machine was used for loading and data acquisition.

The primary purpose of the experiments was to validate the macroscopic load response of the simulations. Hence, all the testing parameters were identical between the experiment and simulation. This includes packing fraction, which was ensured by including the same number of particles in the experiment and simulation. A total of three cases were experimentally tested, as outlined below.
\begin{itemize}
    \item Mono-disperse assembly comprising \SI{3}{mm} particles.
    \item Mono-disperse assembly comprising \SI{5}{mm} particles. 
    \item Segregated assembly comprising \SI{5}{mm} particles in the bottom half and \SI{3}{mm} particles in the top half.
\end{itemize}
The rest of the cases were not experimentally tested as the particle sizes lie within the cases that have been tested. Hence, if the extreme particle sizes yield a good match between the experiment and simulation, similar behaviour can be expected for the intermediate particle sizes.
\section{Triaxial DEM Simulations}
\label{sec: triaxial simulations}
\subsection{Discrete element method}
The discrete element method has been used to simulate the triaxial tests described in~\cref{sec: triaxial experiments}. Hertz-Mindlin contact model~\cite{Hertz1882, Mindlin1949, Mindlin1953} with linear rolling friction has been utilised to model contact between the glass beads. The Hertz-Mindlin contact has four components, as shown below in~\cref{eq:hm_norm,eq:hm_tang,eq:fdn,eq:fdt}.
\begin{equation}   \mathbf{\mathbf{F_\text{n}}} = \frac{4}{3}E^*\sqrt{R^*}\delta^{\frac{3}{2}}_\text{n}
    \label{eq:hm_norm}
\end{equation}
\begin{equation}
    \mathbf{F_\text{t}} = -8G^*\sqrt{R^*\delta_\text{n}}\delta_\text{t}
    \label{eq:hm_tang}
\end{equation}
\begin{equation}
    \mathbf{F^d_\text{n}} = 2*\sqrt{\frac{5}{6}} \left( \frac{\ln e}{\left(\ln e\right)^2 + \pi^2}\right) \sqrt{2 E^* \sqrt{R^* \delta_\text{n}} m^*}\,\mathbf{v^{\text{rel}}_\text{n}}
    \label{eq:fdn}
\end{equation}
\begin{equation}
    \mathbf{F^d_\text{t}} = 2*\sqrt{\frac{5}{6}}\left( \frac{\ln e}{\left(\ln e\right)^2 + \pi^2}\right)\sqrt{8G^* \sqrt{R^* \delta_\text{n}}m^*}\, \mathbf{v^{\text{rel}}_\text{t}}
    \label{eq:fdt}
\end{equation}
where $\mathbf{F_\text{n}}$, $\mathbf{F^d_\text{n}}$, $\mathbf{F_t}$ and $\mathbf{F^d_t}$ corresponds to force due to normal stiffness, normal damping, tangential stiffness and tangential damping, respectively. $E^*$,$R^*$,$G^*$ and $m^*$ represent effective Young's modulus, effective radius of contact, effective shear modulus and effective mass respectively and $\delta_\text{n}$,$\delta_\text{t}$,$v^{rel}_\text{n}$ and $v^{rel}_\text{n}$ represent the normal overlap, tangential overlap, normal relative velocity and tangential relative velocity respectively. The tangential stiffness has a ceiling value dictated by the static friction coefficient in accordance with the Coulomb law of friction.
All the parameters and material parameters defined in the simulations can be seen in \cref{Table: sim_parameters}.

\begin{figure}[htbp!]
     \centering
     \hspace{1 cm}
     \begin{subfigure}{0.15\textwidth}
         \centering
         \includegraphics[width=1\textwidth]{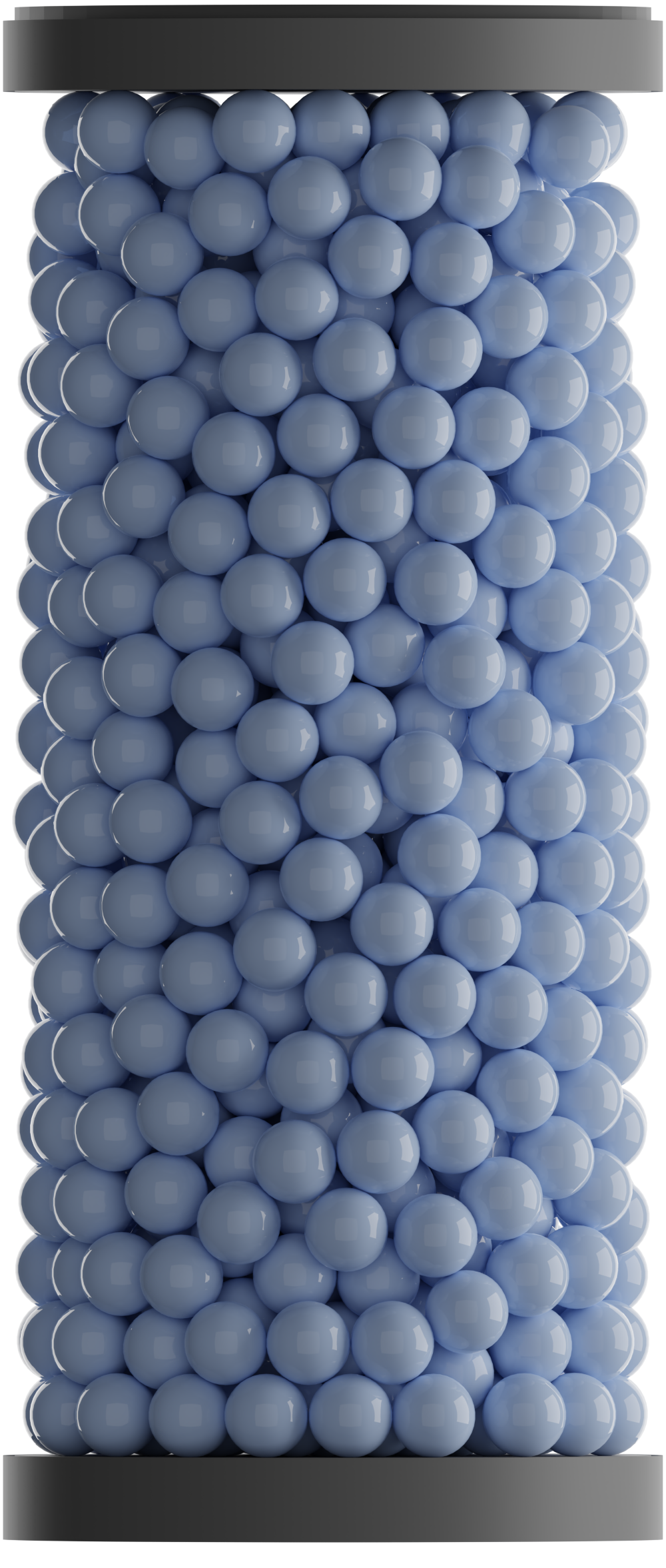}
         \caption{}
         \label{fig:conf_init_5}     
     \end{subfigure}
     \begin{subfigure}{0.15\textwidth}
         \centering
         \includegraphics[width=1\textwidth]{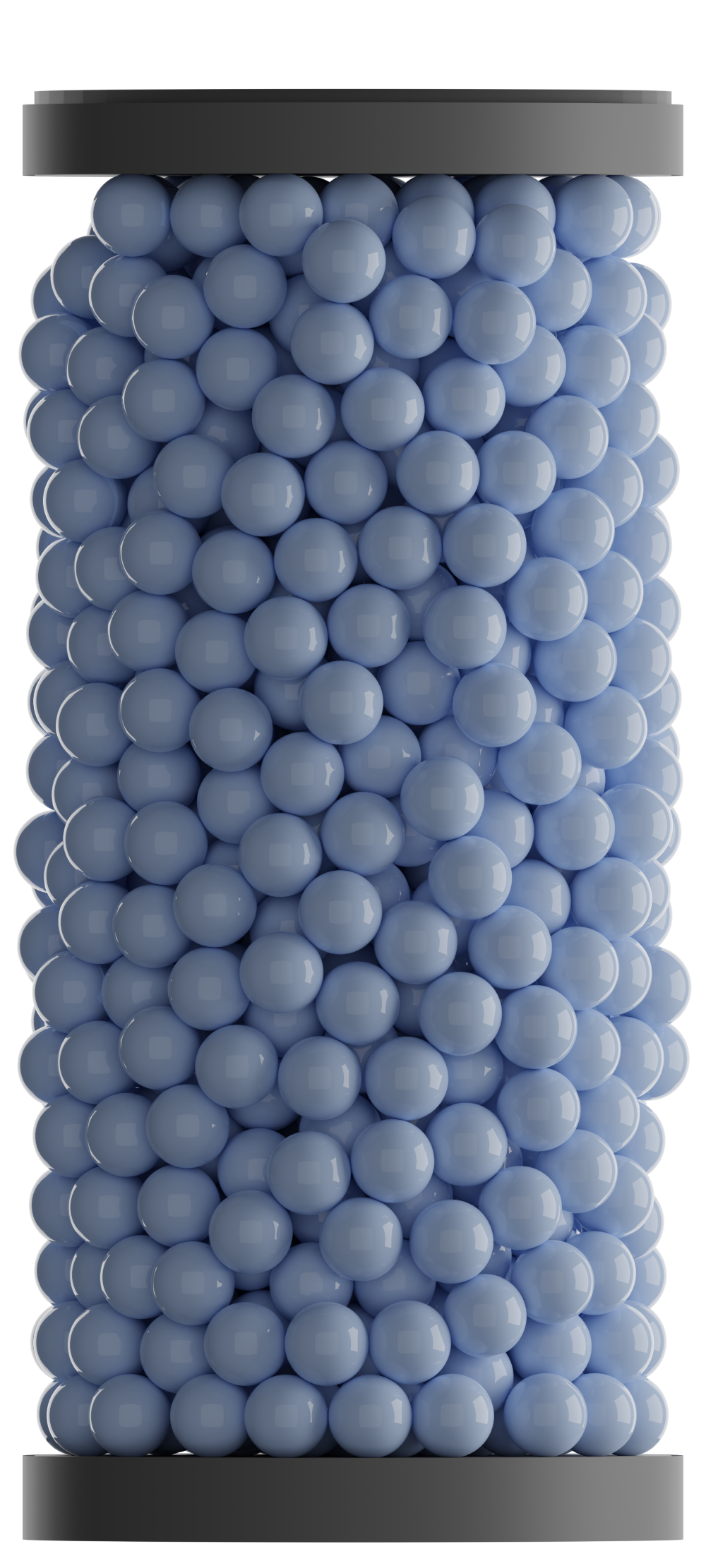}
         \caption{}
         \label{fig:conf_fin_5}     
     \end{subfigure}
     \hspace{1 cm}
     \begin{subfigure}{0.15\textwidth}
         \centering
         \includegraphics[width=1\textwidth]{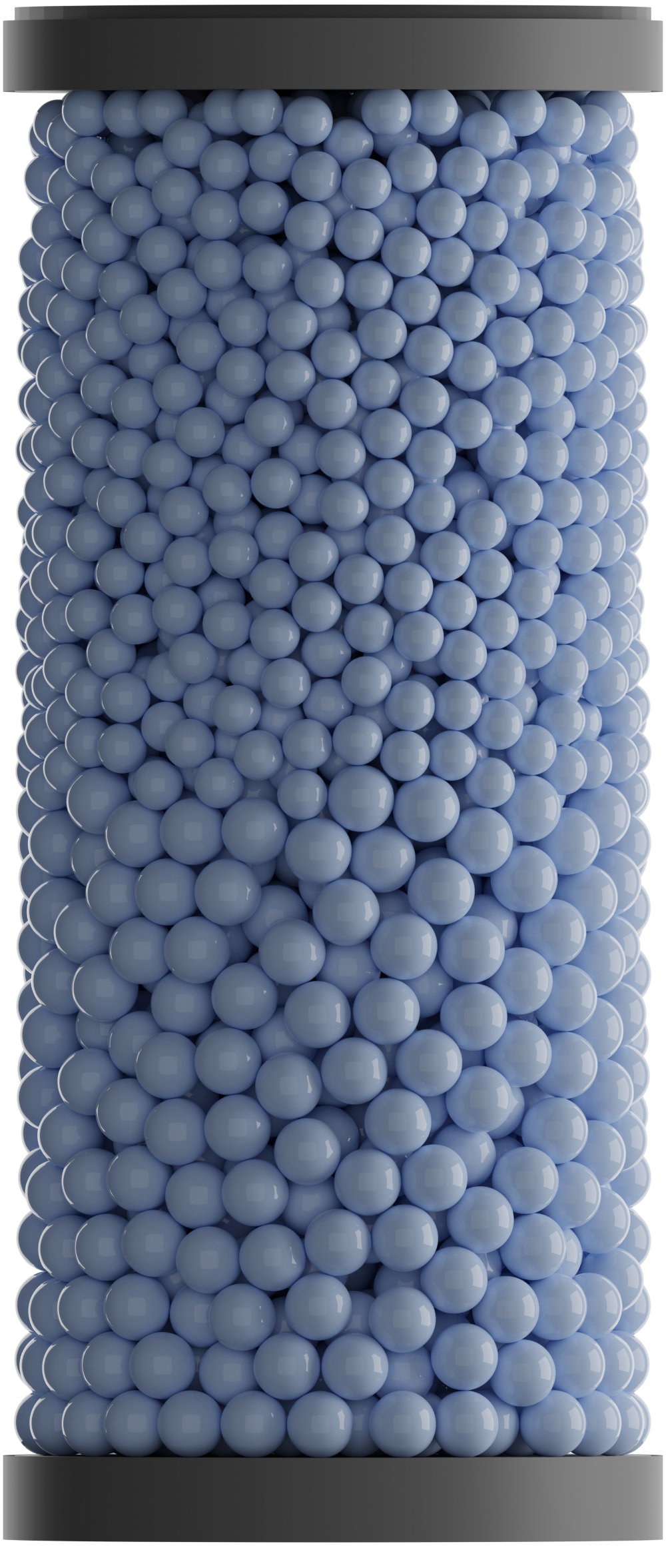}
         \caption{}
         \label{fig:conf_seg_init}
     \end{subfigure}
     \begin{subfigure}{0.15\textwidth}
         \centering
         \includegraphics[width=1\textwidth]{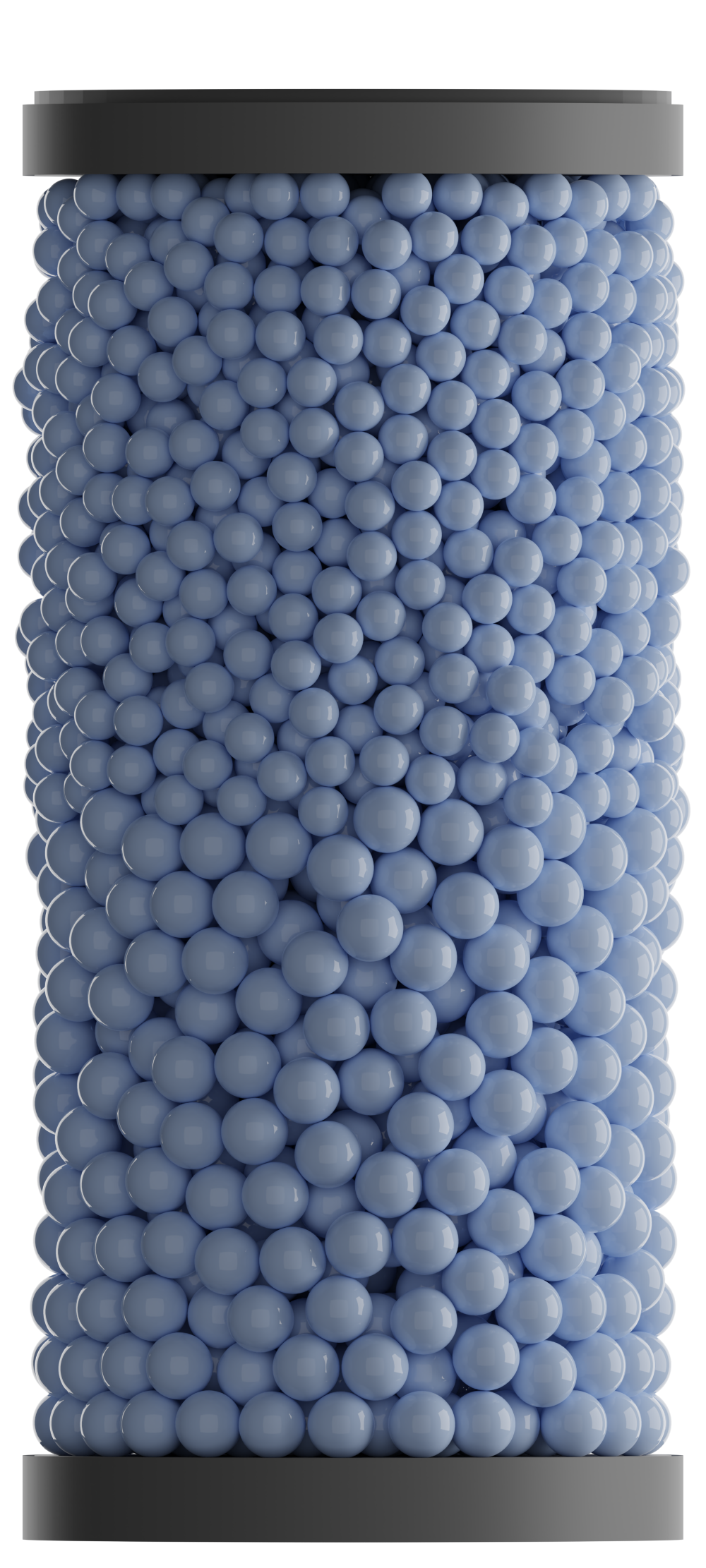}
         \caption{}
         \label{fig:conf_seg_fin}
     \end{subfigure}
        \caption{(a) Initial and (b) final configuration of the mono-disperse specimen comprising of \SI{5}{mm} particles and (c) Initial and (d) final configuration of the segregated specimen comprising of \SI{5}{mm} in the bottom half and \SI{3}{mm} particles in the top half.}
        \label{fig:assemblies}
\end{figure}

The rolling friction model used in the simulations can be seen in \cref{eq:rolling friction}, where $\mathbf{\tau}$ is the torque applied on the particle to resist rolling. $\mathbf{\tau}$ is applied in the direction opposite to the angular velocity vector of the particle in consideration. $R_i$ is the distance between the centre of the particle and the point of application of the torque. This point is the midpoint of the line segment joining the centres of the particles in contact. $\omega_i$ is the unit vector along the direction of angular velocity.
\begin{equation}
    \mathbf{\tau} = - \mu_\text{r} \mathbf{F_\text{n}} R_i \omega_i
    \label{eq:rolling friction}
\end{equation}
The glass particles are first consolidated in a cylindrical container of radius \SI{38}{\milli\meter} and height \SI{80}{\milli\meter}. Subsequently, membrane particles~(\cref{membrane_ss}) are generated and fixed between the top and bottom cylindrical platens, similar to the experiment. A confinement force is then gradually applied~(\cref{confinment_ss}) by applying a central force on the membrane particles. Upon confinement, the top platen is then moved down at a rate of \SI{5}{mm/min} up to a total displacement of \SI{10}{mm}. A total of five specimens are simulated. Two mono-disperse specimens comprising of \SI{3}{mm} and \SI{5}{mm} diameter particles, respectively and three segregated specimens comprising of \SI{3}{mm} diameter particles in the bottom half and \SI{4}{mm}, \SI{4.5}{mm} and \SI{5}{mm}  diameter particles in the top half respectively. The initial and final configuration of the one mono-disperse specimen and one segregated specimen can be seen in \cref{fig:assemblies}. Altair EDEM\texttrademark has been used to run the DEM simulations presented in this article. Custom particle-generator and custom body force models have been incorporated into the program to simulate the membrane and confinement.
\begin{table}[htbp!]
\centering
\caption{Simulation parameters}
\label{Table: sim_parameters}
\begin{tabular}{ll|ll}
\hline
\textbf{Property} & \textbf{Value} & \textbf{Property} & \textbf{Value} \\ \hline
\textbf{Young's modulus} &  & \textbf{Static coefficient of friction} &  \\
\, glass particles & \SI{70}{\giga\pascal} & \, glass-glass & 0.5 \\
\, membrane particles & \SI{0.25}{\giga\pascal} & \, glass-membrane & 0.1 \\
\, loading plate & \SI{0.25}{\giga\pascal} & \, glass-loading plate & 0.1 \\ \hline
\textbf{Poisson's ratio} &  & \textbf{Coefficient of Restitution} &  \\
\, glass particles & 0.22 & \, glass-glass & 0.7 \\
\, membrane particles & 0.25 & \, glass-membrane & 0.5 \\
\, loading plate & 0.25 & \, glass-loading plate & 0.5 \\ \hline
\textbf{Density} &  & \textbf{Coefficient of rolling friction} &  \\
\, glass particles & \SI{2500}{\kilo\gram\per\meter^3} & \, glass-glass & 0.01 \\
\, membrane particles & \SI{1100}{\kilo\gram\per\meter^3} & \, glass-membrane & 0.01 \\
 &  & \, glass-loading plate & 0.01 \\ \hline
\end{tabular}%
\end{table}

\subsection{Simulating the membrane}
\label{membrane_ss}
\begin{figure}[htb]
\begin{center}
\includegraphics[width=7 cm]{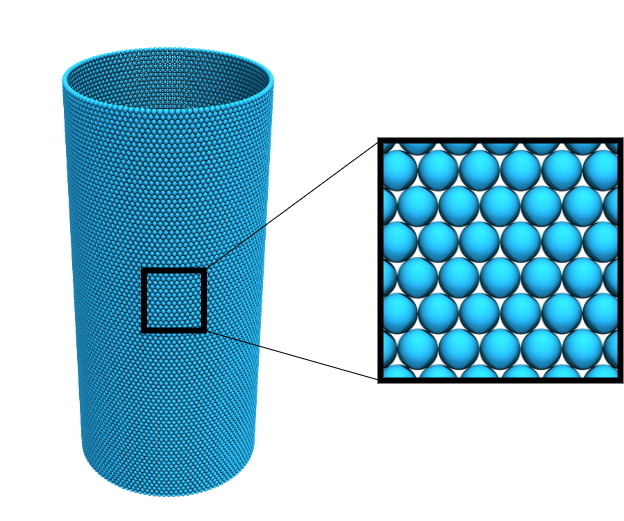}
\caption{Visualisation of the hexagonal close packing arrangement of membrane particles in a cylinder that will blanket the glass beads.}
\label{Fig:membrane_cp}
\end{center}
\end{figure}

\begin{figure}[htbp!]
    \centering\includegraphics[width=0.5\linewidth]{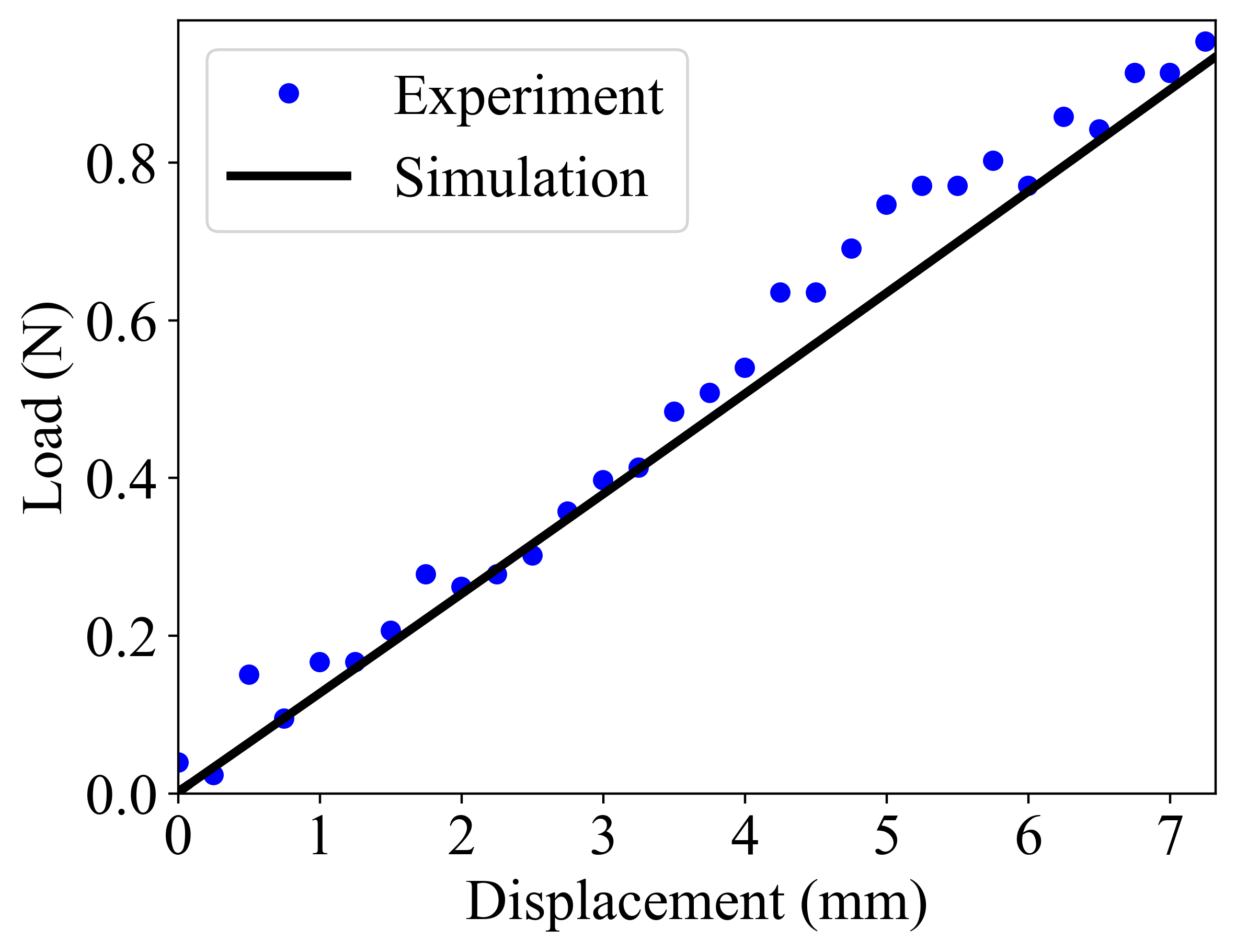}
    \caption{Comparison between experiment and simulation for uni-axial loading of a $\SI{8}{mm}\times\SI{4}{mm}$ strip of membrane.}
    \label{fig:membrane_test}
\end{figure}

The membrane used in triaxial tests is simulated by a cylindrical layer of bonded membrane particles around the glass particles. The radius of the membrane particles is chosen as \SI{0.5}{mm}, which is a third of the smallest particle in the specimen. The properties of the membrane particles can be seen in \cref{Table: sim_parameters}. The membrane particles are arranged in a hexagonal close packing structure to minimise inter-particle gaps, as seen in~\cref{Fig:membrane_cp}.
Immediately after the generation of membrane particles, bonds are generated between particles in contact according to a bond formation radius, which was set at 150\% the physical radius of the particle. This allows for the bonding of the immediate neighbours amongst the particles.The bond normal stiffness is set to \SI{300}{\newton\per\meter} and a tangential stiffness of \SI{150}{\newton\per\meter}. These values have been obtained by performing uni-axial tensile loading of a $\SI{8}{mm}\times\SI{4}{mm}$ strip of membrane. This is followed by replicating the tensile test in DEM with our particle membrane and calibrating the parameters to match the macroscopic stiffness of the strip. The comparison between the experiment and simulation can be seen in~\cref{fig:membrane_test}. 
\subsection{Simulating confinement}
\label{confinment_ss}
Confinement has been modelled by applying a radially varying central force on the membrane particles. This implementation of confinement is only suitable for axisymmetric modes of membrane deformation. Such axisymmetric deformation is seen only when the top platen is non-rotatable; hence, only those cases have been simulated. This confinement force has been derived, assuming each membrane particle represents a small area of the membrane. These area elements can be assumed to be hexagonal, but the shape is irrelevant to the confinement force. The total confinement force~($F_c$) due to the desired confinement pressure~($P_c$) is calculated~(\cref{total_force}), from which the initial force~($F^\text{init}_p$) on each particle is calculated~(\cref{Force_per_particle_init}). Assuming the membrane expands axisymmetrically, the area represented by each element and, subsequently, the force on each particle will increase linearly with an increase in the radial position of the particle from the central axis. Also, the reduction in the height of the membrane is incorporated by proportionally reducing the height represented by each particle proportionally. This results in a proportional reduction of area and, hence, confinement force. The position-dependent force at a time t~($F_{\text{p}}^\text{t}$) on each particle is therefore given by \cref{Force_per_particle} below.
\begin{equation}
    F_{\text{c}} = P_{\text{c}}\times 2\pi r_\text{s}^{\text{init}} h_\text{s}^{\text{init}}
    \label{total_force}
\end{equation}
\begin{equation}
F^{\text{init}}_{\text{p}} = \frac{F_\text{c}}{N_{\text{p}}}    \label{Force_per_particle_init}
\end{equation}
\begin{equation}
    F_{\text{p}}^\text{t} = \frac{h_\text{s}^{\text{init}} - \Delta \text{h}_\text{t}}{h_\text{s}^{\text{init}}}\times\frac{r_\text{t}}{r_\text{s}^{\text{init}}} \times F^{\text{init}}_{\text{p}}
  \label{Force_per_particle}
\end{equation}
where $r_\text{s}^{\text{init}}$ and $r_\text{t}$ are the initial and current radial distance of the membrane particle from the axis of the membrane cylinder, respectively. $h_\text{s}^{\text{init}}$ and $\Delta \text{h}_\text{t}$ are the initial height and the change in the height of the specimen at time $t$, respectively. $N_\text{p}$ corresponds to the total number of membrane particles.

\begin{figure}[htb]
     \centering
     \begin{subfigure}{0.3\textwidth}
         \centering\includegraphics[width=0.8\textwidth]{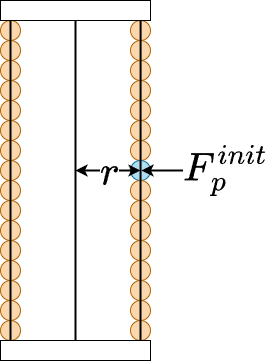}
         \caption{}
         \label{fig:con_1}
     \end{subfigure}
     \begin{subfigure}{0.3\textwidth}
         \centering\includegraphics[width=0.8\textwidth]{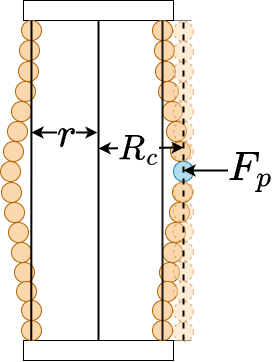}
        \caption{}
         \label{fig:con_2}
     \end{subfigure}
        \caption{(a)Initial and (b) Final configuration of the membrane particles indicating the initial radius $r$ and final radius $R_c$ from the central axis for one of the particles.}
        \label{fig:conexp}
\end{figure}

It should be noted that sudden application of this force will cause instabilities in the system. Hence, the force is gradually applied over a time of \SI{1}{\second}. The rationale behind this model of force application is as follows. It can be assumed that a membrane particle deforming from the initial value of $r_\text{s}^{\text{init}}$ becomes part of a new imaginary membrane of radius equal to the final value of $r_\text{t}$. This can be seen in \cref{fig:conexp}, as the membrane particle highlighted in blue moves laterally, it is assumed to be part of an imaginary membrane with a radius $r_\text{t}$. The surface area of the imaginary membrane would be scaled by ${r_\text{t}}/{r_\text{s}^{\text{init}}}$ when compared to the original membrane. The force due to the confinement pressure would be directly proportional to the surface area; hence it is scaled by ${r_\text{t}}/{r_\text{s}^{\text{init}}}$ as seen in~\cref{Force_per_particle}.

This approach to modelling confinement has certain shortcomings. Firstly, the simulations are limited to axisymmetric cases, which means cases with rotating loading platens cannot be simulated. Apart from that, the assumptions made about the deformation method of the area element represented by the membrane particles are approximate and assume ideal deformation cases. The following subsection deals with validating the confinement force applied to establish its accuracy.
\subsection{Validating confinement}
\label{confinment_validation_ss}
The confinement model presented above does not involve any feedback mechanism or area calculation. This means that the confinement pressure is not controlled during the simulation. To ensure the reliability of the confinement model, a retrospective post-processing procedure to measure the confinement pressure is proposed. The first step in computing the confinement pressure is obtaining the required simulation data. We need the contact forces transmitted between the membrane particles and the specimen particles and the positions of points of contact to reconstruct the membrane as a mesh. The next step is to arrange rectangles along the membrane. The position of the rectangles is represented in polar coordinates. The $z$ and $\theta$ components of the position of the rectangles are fixed to span the whole membrane. The elements are angled such that their surface normals are always radially outward. To determine the radial component of the position of a rectangle, the radial positions of all the contact points whose $z$ and $\theta$ components fall within the limits of the rectangle are aggregated. This ensures the array of rectangles tile the entire membrane. Subsequently, the centres of each of the rectangles are connected to form a mesh composed of quadrilateral elements. For a specific element in this mesh, the four nodes' radial and $z$ components are used as limits to find the contact points that lie `within' the element. The contact force vectors associated with each contact point are added to find the total force transmitted through the element. This cumulative force vector and the area vector of the element are used to calculate the pressure associated with the element.

\begin{figure}[htbp!]
     \centering
     \begin{subfigure}{0.208\textwidth}
         \centering\includegraphics[width=\textwidth]{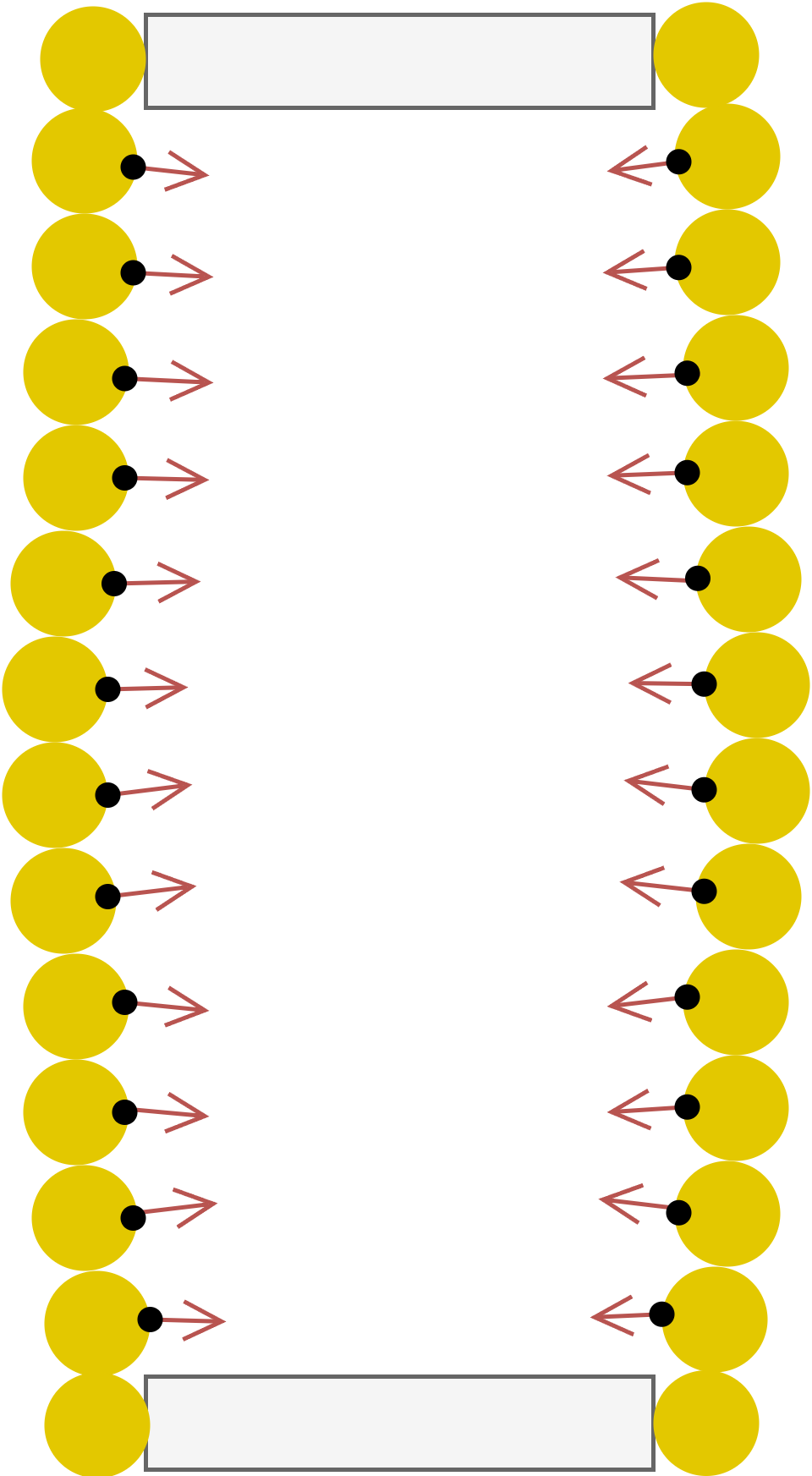}
         \caption{}
         \label{subfig:pt1}
     \end{subfigure}
     \begin{subfigure}{0.17\textwidth}
         \centering\includegraphics[width=\textwidth]{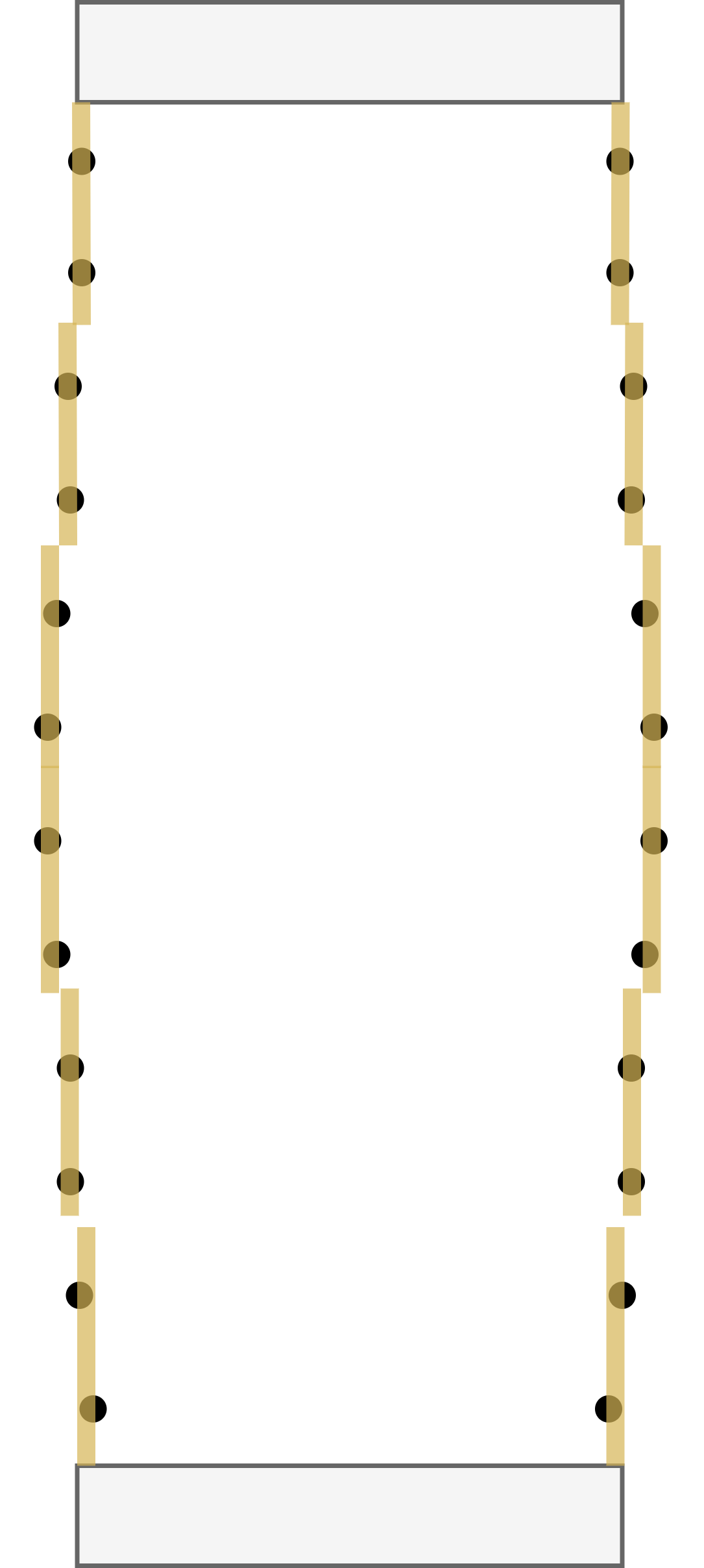}
         \caption{}
         \label{subfig:pt2}
     \end{subfigure}
     \begin{subfigure}{0.17\textwidth}
         \centering\includegraphics[width=\textwidth]{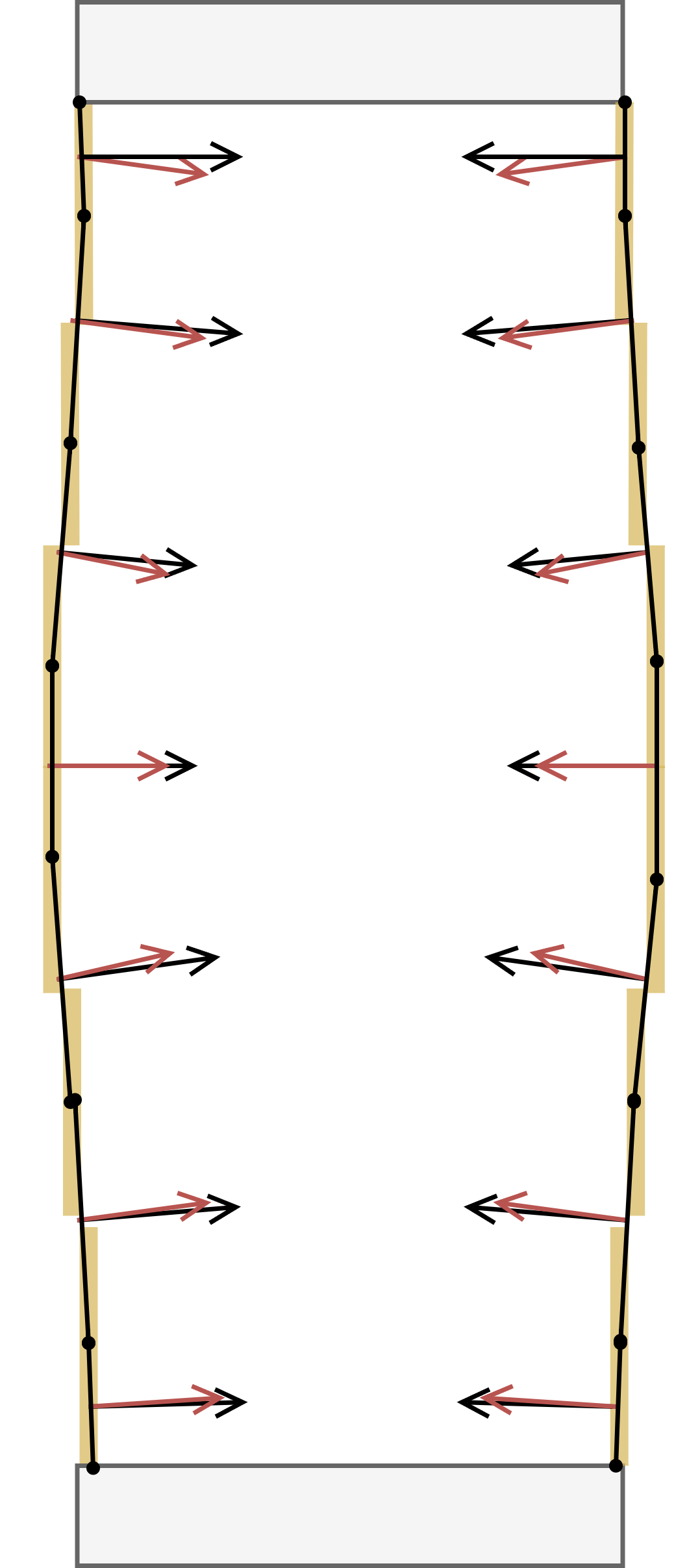}
         \caption{}
         \label{subfig:pt3}
     \end{subfigure}
        \caption{(a) The membrane particles and contact force vectors between the membrane particle and specimen particles along with the contact points~(b) Placing the rectangles~(lines in 2D) along the membrane.~(c) Creating the final mesh from the centres of the rectangles~(lines) shown in (b), along with the corresponding normal vectors in black and contact force vectors in red.}
    \label{fig:confinement}
\end{figure}

An illustrative example of the methodology explained above in 2D can be seen in~\cref{fig:confinement}. \cref{subfig:pt1} shows the membrane particles and the force the particle is applying on the specimen. Tiling the membrane with rectangles is seen in~\cref{subfig:pt2}, except the rectangles are replaced by vertical lines. The centres of the lines are joined to generate a mesh, as seen in~\cref{subfig:pt3}, wherein the area vectors, along with the cumulative force vector associated with each mesh element, are also known. This pressure is averaged over all the elements of the mesh to obtain the average confinement pressure being applied.

\begin{figure}[htbp!]
    \centering\includegraphics[width=0.5\linewidth]{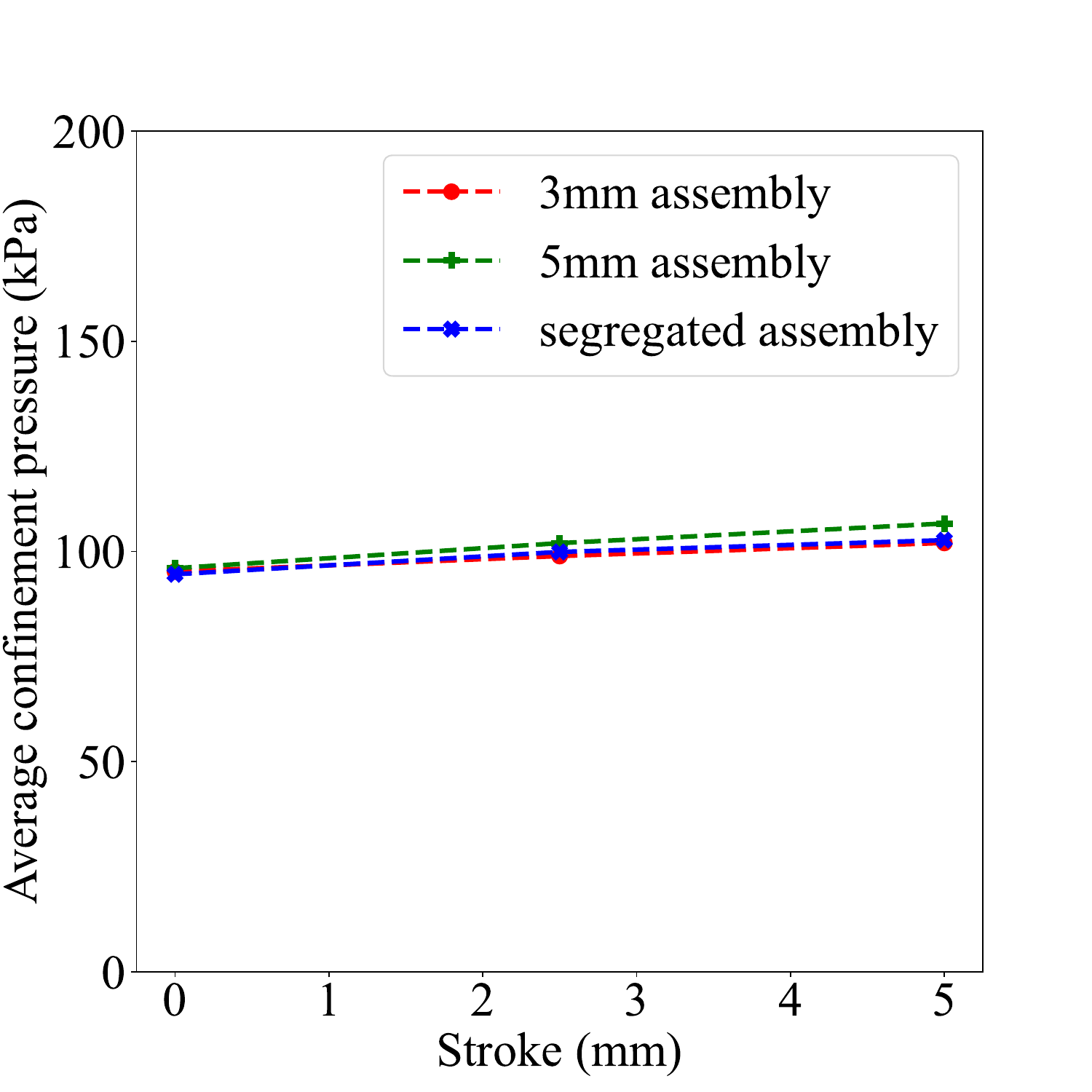}
    \caption{Evolution of average confinement pressure for the mono-disperse assemblies and the segregated assembly comprising the \SI{3}{mm} and \SI{5}{mm} particles.}
    \label{fig:pressure_test}
\end{figure}

The average pressure in the initial, middle and final configurations of the simulations can be seen in~\cref{fig:pressure_test}. Both the mono-disperse cases, along with the segregated case comprising \SI{3}{mm} and \SI{5}{mm} particles, are tested. Other segregated cases~(\SI{3}{mm} with \SI{4}{mm} and \SI{3}{mm} with \SI{4.5}{mm}) are not tested as any discrepancy arising in the intermediary particle sizes would also be observed in the extreme particle sizes that have been tested. The desired confinement pressure is \SI{100}{kPa}. It can be seen that the confinement pressure has a very small positive gradient, but within the \SI{5}{mm} stroke applied, which corresponds to a macroscopic strain of 0.0625, the difference is less than 5\%. This difference is sufficiently small enough to ensure the accuracy of simulations.
\section{Results and discussions}
\label{sec:r_and_d}
\subsection{Load response}
\begin{figure}[htb] 
     \centering
     \begin{subfigure}{0.4\textwidth}
         \centering\includegraphics[width=\textwidth]{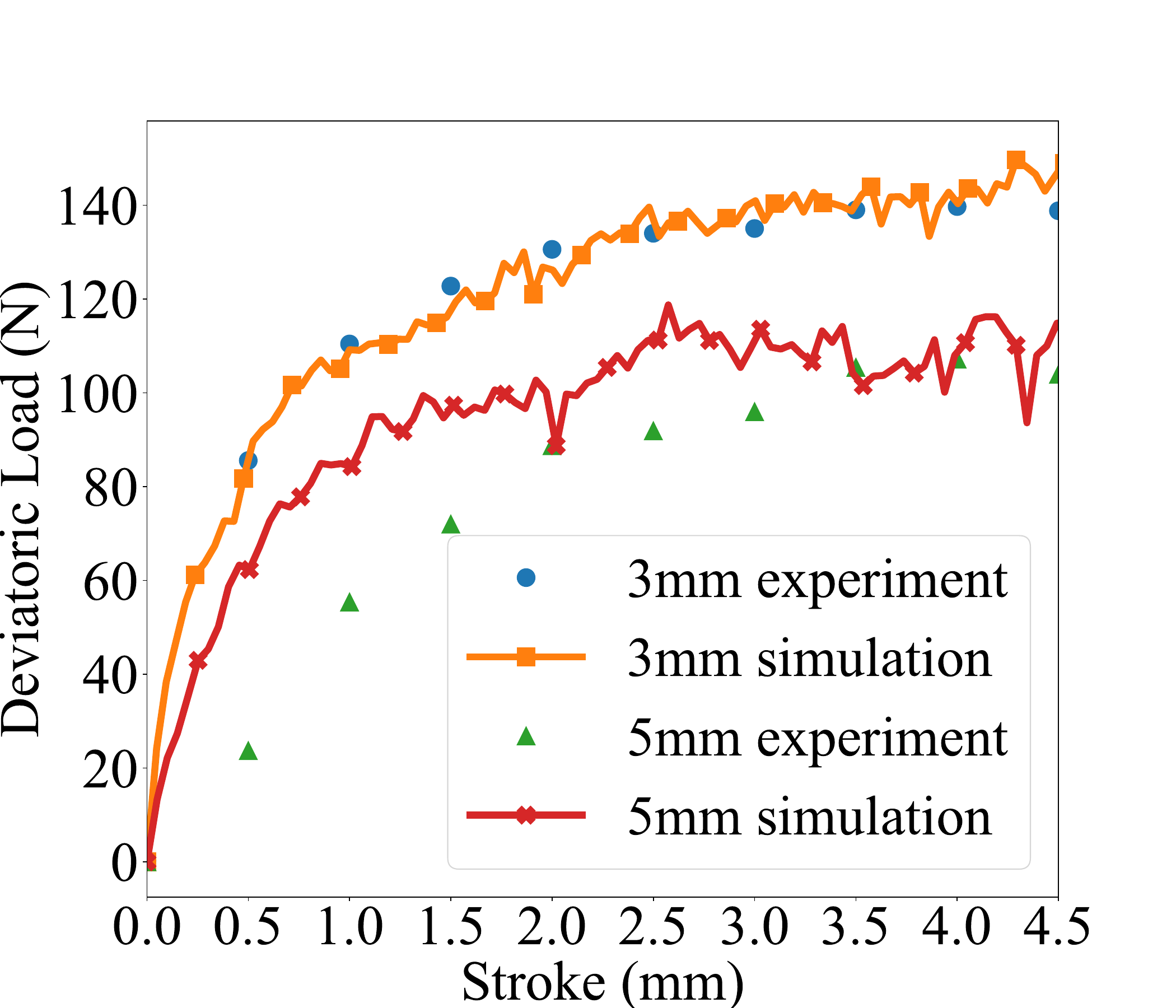}
         \caption{}
         \label{fig:exp_lvd}
     \end{subfigure}
     \begin{subfigure}{0.4\textwidth}
         \centering\includegraphics[width=\textwidth]{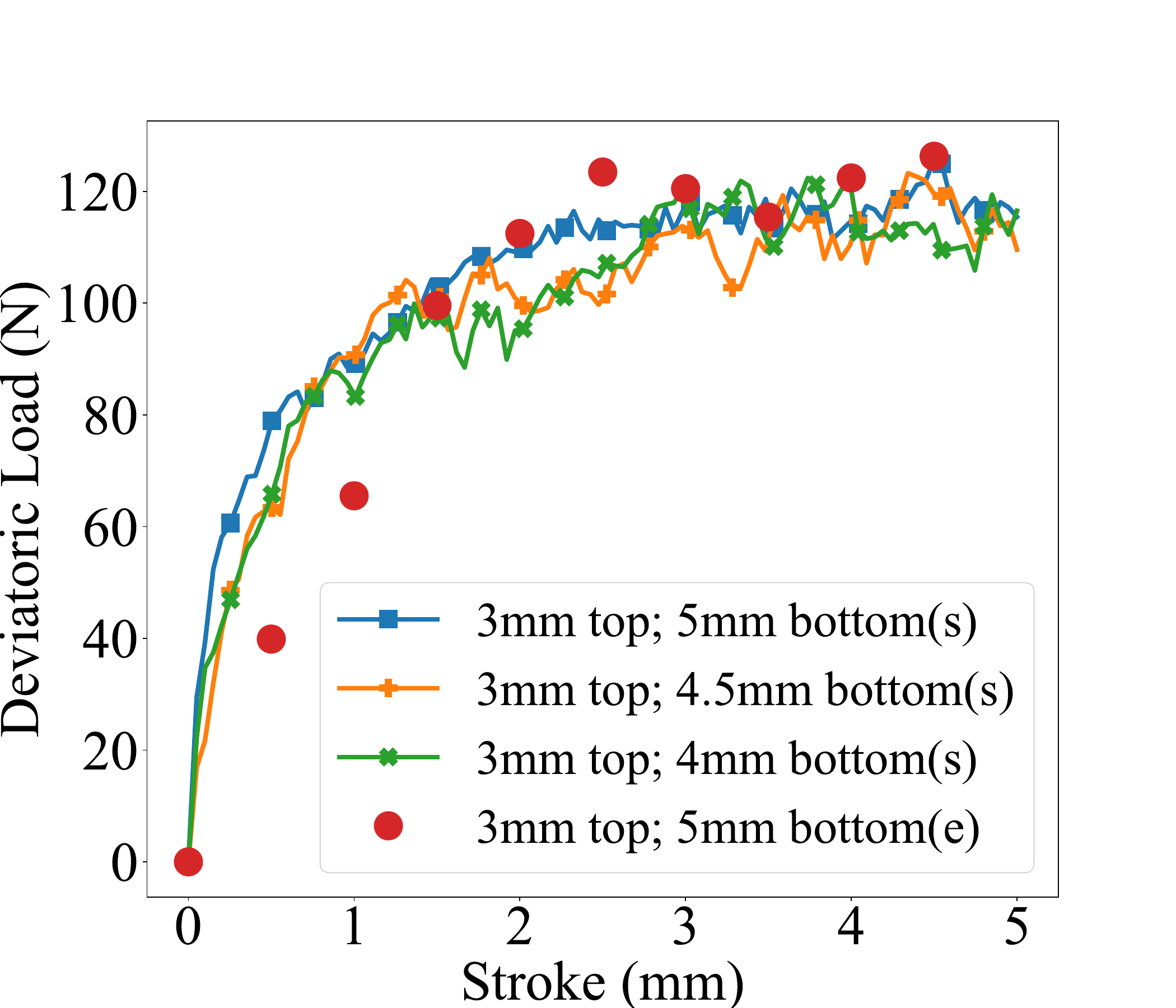}
         \caption{}
         \label{fig:sim_lvd}
     \end{subfigure}
        \caption{(a)The deviatoric load response for the two mono-disperse assemblies comprising experiment and simulation. (b)The deviatoric load response for the three segregated assemblies, comprising experimental response of the assembly comprising~\SI{3}{mm} and~\SI{5}{mm} particles.}
        \label{fig:lvd}
\end{figure}

We first look at the load response of the simulations and, wherever available, the corresponding experiments. This mainly aims to validate our simulation with the corresponding experiment. A good match between the simulation and experiment means the contact parameters that we have chosen are closer to the realistic values.~\cref{fig:exp_lvd} shows the response for the~\SI{3}{mm} mono-disperse assembly and the~\SI{5}{mm} mono-disperse assembly. We see a good agreement between the experiments and the simulation for both cases. We see that the~\SI{3}{mm} assembly exhibits a higher load-bearing capacity than the ~\SI{5}{mm} assembly, seen both in experiment and simulation. ~\cref{fig:sim_lvd} shows the comparison of the load response obtained for the segregated assemblies. This includes the experimental result for the segregated assembly comprising the \SI{3}{mm} and \SI{5}{mm} particles, which again is in good agreement with the simulation. All the segregated assemblies have similar load responses. When compared to the mono-dispersed assemblies, it is seen that the segregated assemblies have a load response closer to the \SI{5}{mm} particles. 

The main aim of this comparison is to validate the load response of the simulations and experiments. Good agreement between the simulation and experiment is a decent indicator towards accurate inter-particular contact behaviour. Especially since it has already been established that the correct confinement is being applied, this, in turn, translates to accurate rheology during loading. It should also be noted that we have experimentally verified cases with extreme particle sizes~(\SI{3}{mm} and \SI{5}{mm}). If the discrepancy in applying the confinement pressure is not present for extreme particle sizes in a segregated assembly, such a discrepancy would not be present for all the intermediate particle sizes.
\subsection{Strain}
\begin{figure}[htb]
    \centering   \includegraphics[width=1\textwidth]{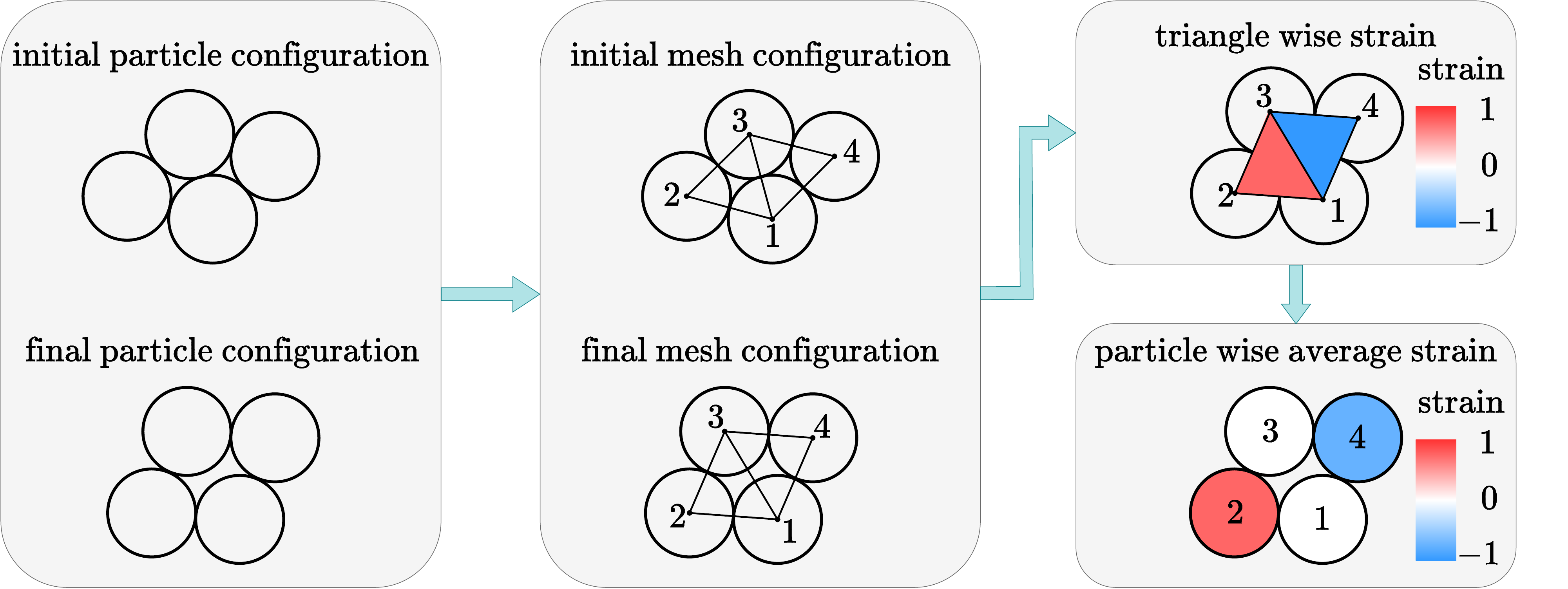}
    \caption{A simplified workflow for calculating an arbitrary component of the strain tensor for a four-particle system}
    \label{fig:sf}
\end{figure}

The particle position data is used to calculate a `per-particle average' strain data, and the methodology is described in~\cref{fig:sf}. First, Delaunay triangulation is used to mesh the domain using particle centres as nodes. This results in a mesh comprising tetrahedral elements. Some of the elements are deleted if they fail based on a maximum aspect ratio criterion to avoid erroneous strain calculations. The value of the maximum allowable aspect ratio was set to 5. Subsequently, mesh deformation is tracked via nodal displacement obtained from DEM data. From this, constant strain tetrahedron formulation is used to calculate element-wise strain using the Green strain formulation, which can be seen in \cref{gs}. Green strain is used because the strain is calculated with respect to the initial/original configuration. Hence, the small-strain approximation cannot be made. The per-particle average strain tensor is obtained by averaging the components of the strain tensor, which belong to all the elements for which the particle is a node. More details about strain calculation used here can be seen in this work \cite{o2011particulate}.
\begin{equation}
    E_{ij} = \frac{1}{2}\left(u_{i,j} + u_{j,i} + u_{k,i}u_{k,j}\right)
    \label{gs}
\end{equation}
For the purpose of visual representation, the strain data is interpolated over a mesh constructed by Delaunay triangulation of the particle centres. This triangulation is only for visual representation and is separate, albeit similar to the mesh used for strain calculation.  A $y$-normal plane passing through the centre of the assembly is used to slice the interpolated data, and the $zx$-component of the strain is shown on this slice.

\begin{figure}[htb]
     \centering
     \begin{subfigure}{0.32\textwidth}
         \centering\includegraphics[width=1\textwidth]{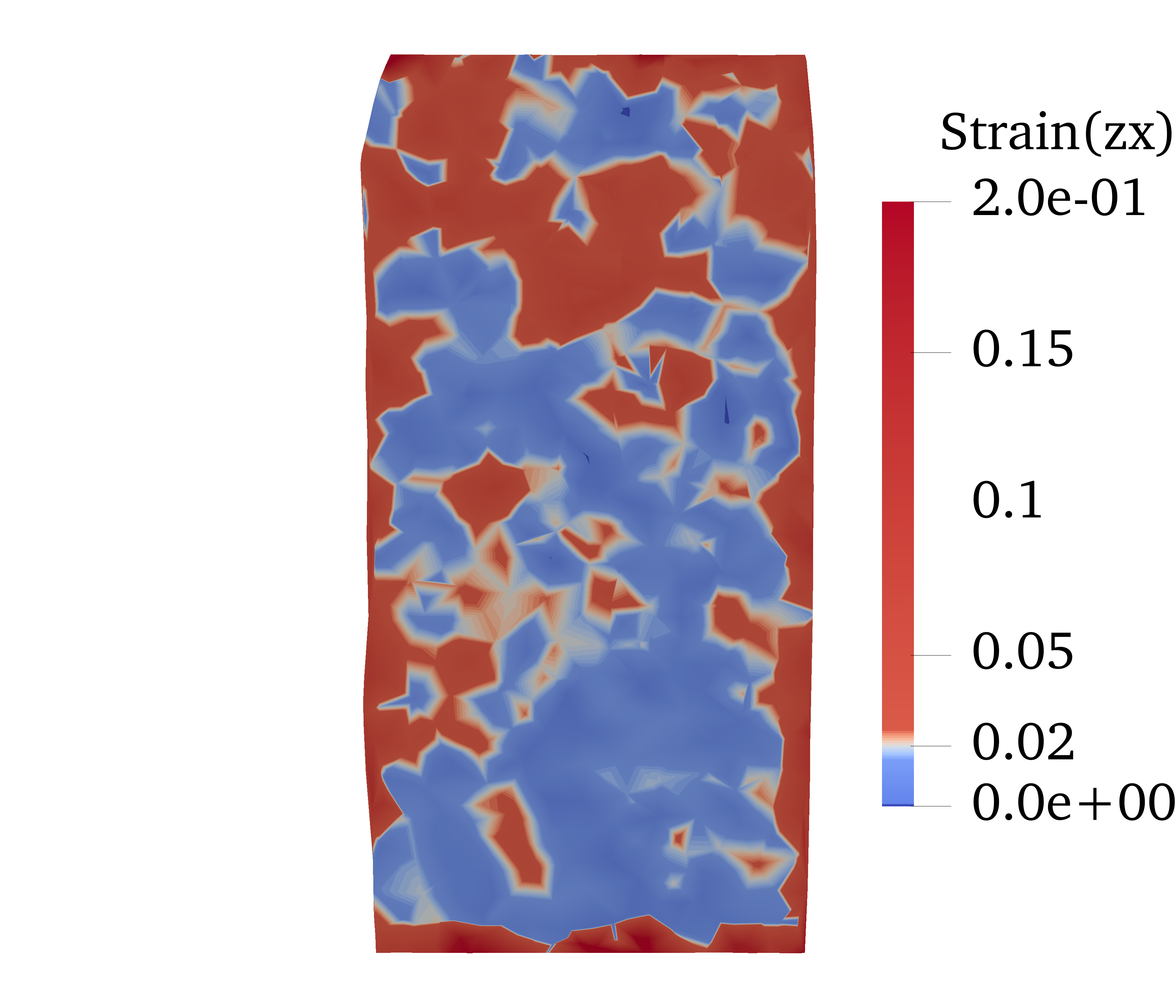}
         \caption{\SI{3}{mm} monodisperse assembly}
         \label{fig:3mm_gzx}
     \end{subfigure}
     \begin{subfigure}{0.32\textwidth}
         \centering
         \includegraphics[width=1\textwidth]{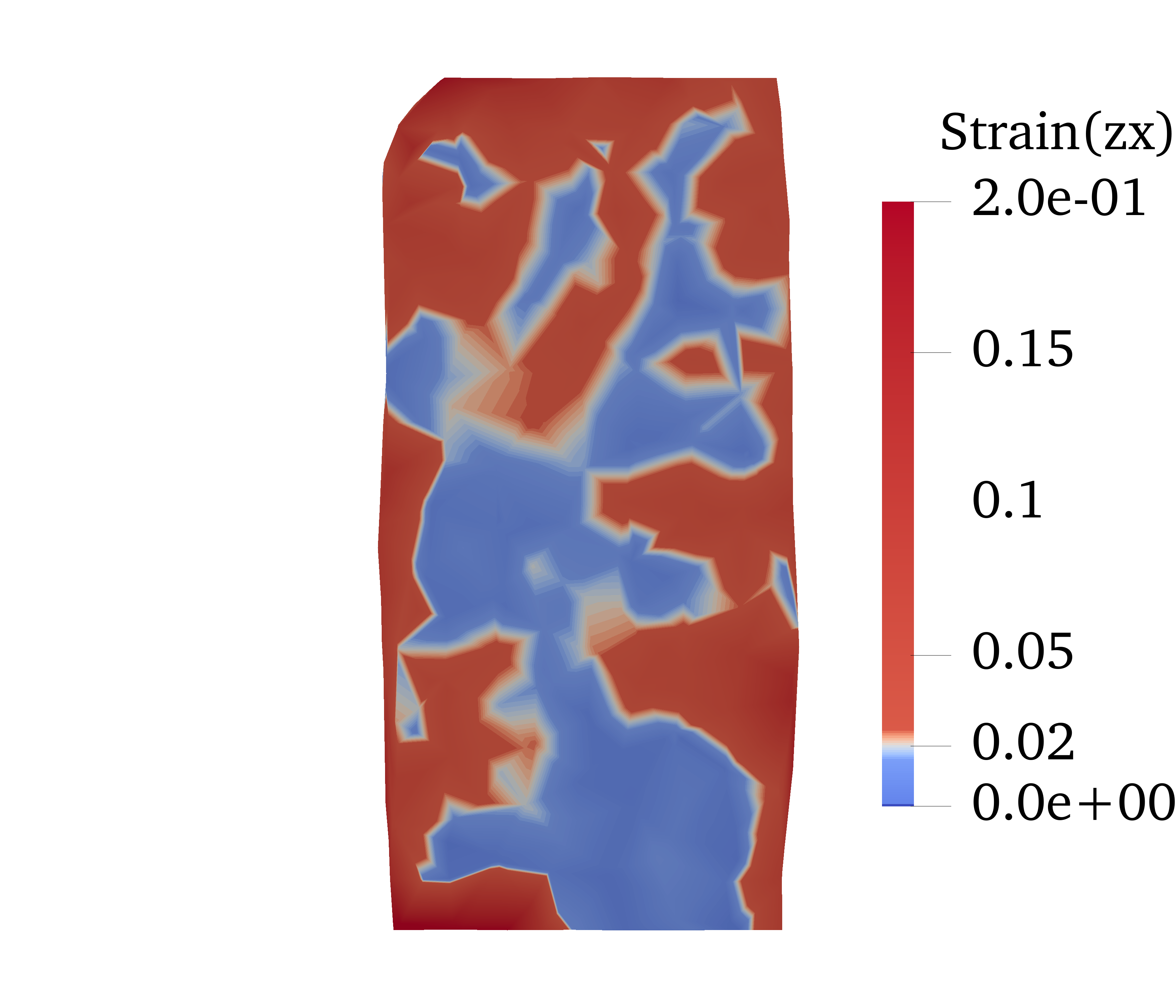}
         \caption{\SI{5}{mm} monodisperse assembly}
         \label{fig:5mm_gzx}
     \end{subfigure}
        \caption{For the mono-disperse assemblies $zx$ component of the Green strain tensor along a plane passing through the centre of the specimen and perpendicular to the $y$-axis.}
        \label{fig:mono_sf}
\end{figure}

\begin{figure}[htb]
     \centering
     \begin{subfigure}{0.32\textwidth}
         \centering\includegraphics[width=1\textwidth]{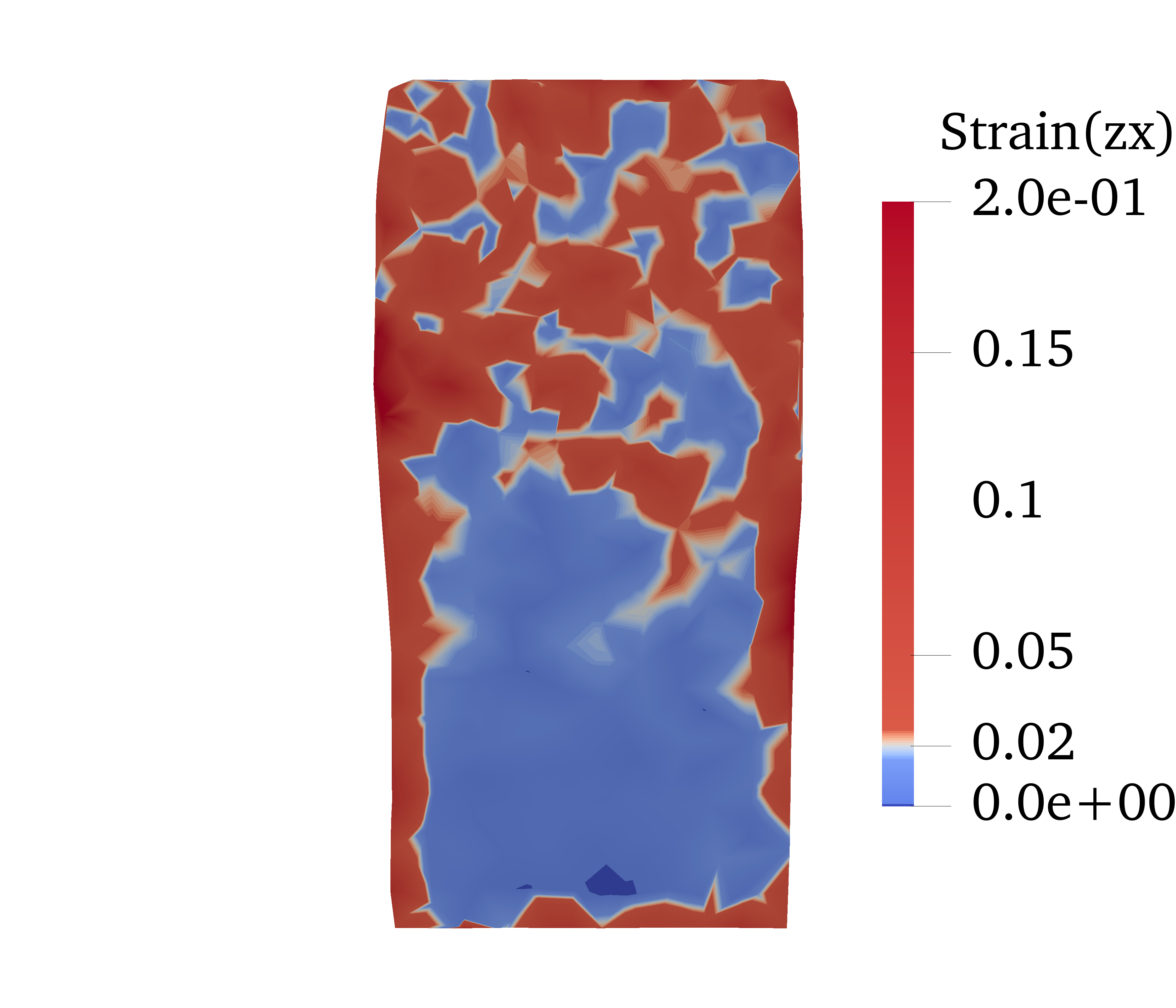}
         \caption{\SI{4}{mm} particles in bottom half and  \SI{3}{mm} particles in top half}
         \label{fig:34mm_gzx}
     \end{subfigure}
     \begin{subfigure}{0.32\textwidth}
         \centering
         \includegraphics[width=1\textwidth]{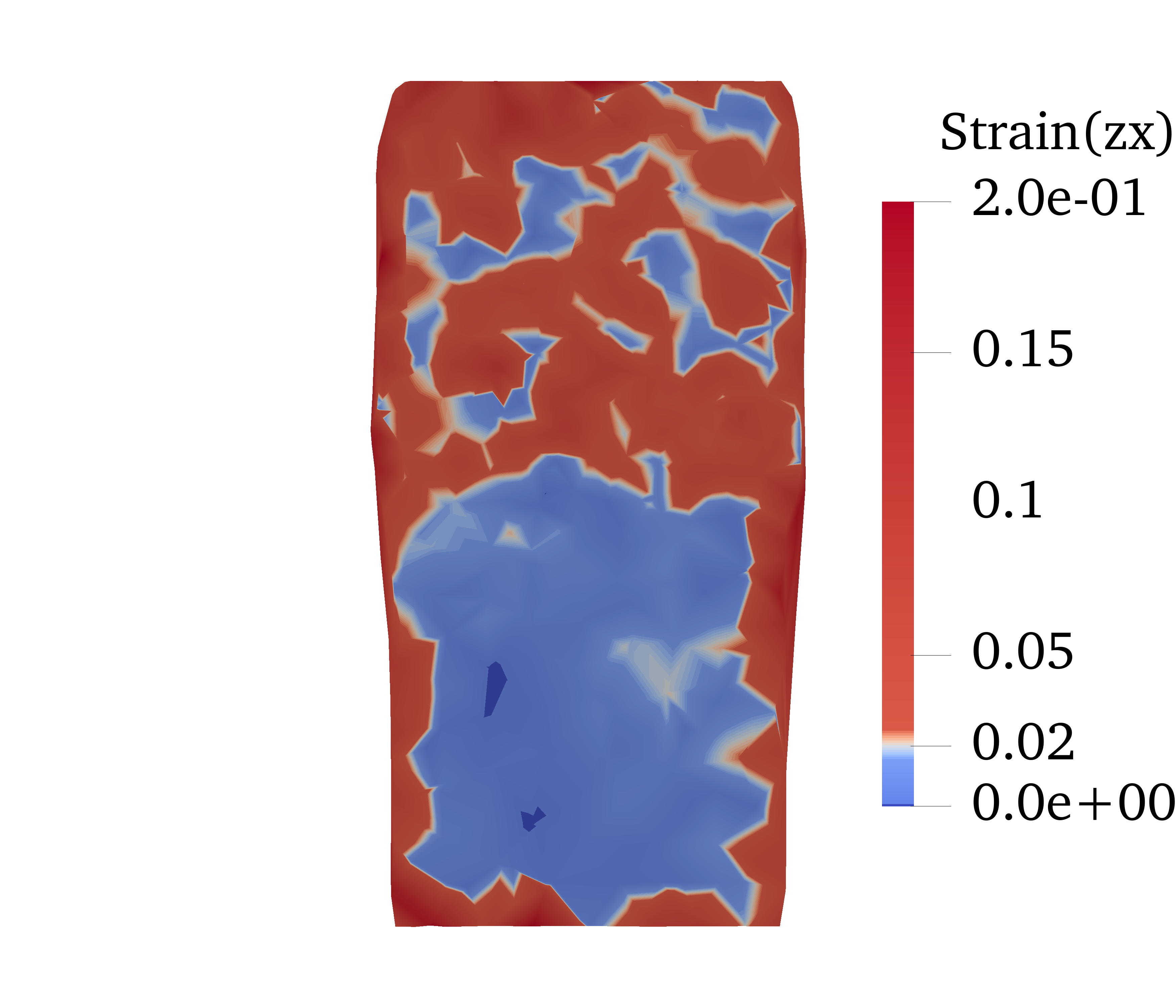}
         \caption{\SI{4.5}{mm} particles in bottom half and  \SI{3}{mm} particles in top half}
         \label{fig:34o5mm_gzx}
     \end{subfigure}
     \begin{subfigure}{0.32\textwidth}
         \centering
         \includegraphics[width=1\textwidth]{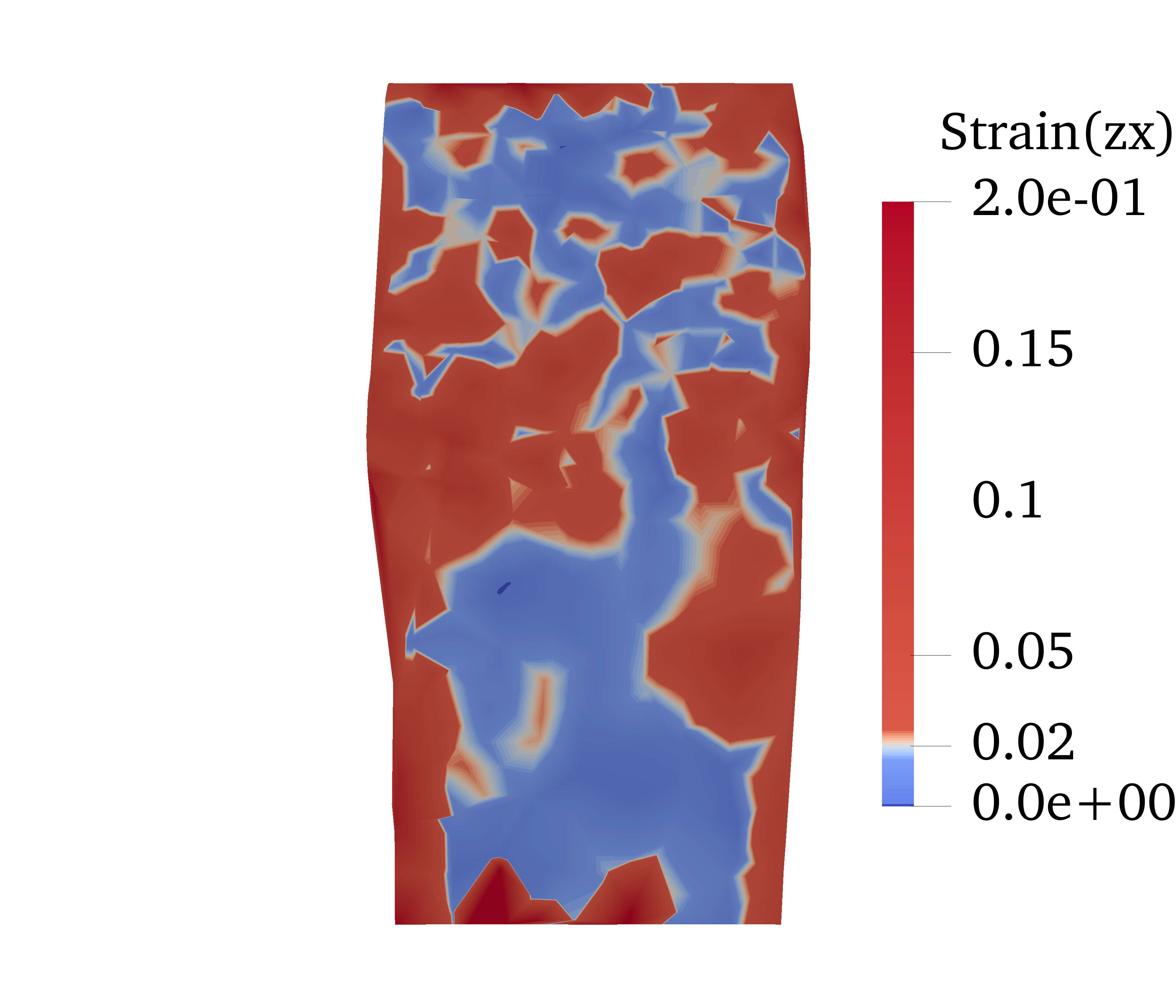}
         \caption{\SI{5}{mm} particles in bottom half and  \SI{3}{mm} particles in top half}
         \label{fig:35_gyz}     
     \end{subfigure}
        \caption{For the segregated assemblies $zx$ component of the Green strain tensor along a plane passing through the centre of the specimen and perpendicular to the y-axis.}
        \label{fig:seg_sf}
\end{figure}

\cref{fig:mono_sf} shows the $zx$ component of Green strain tensor distribution for the \SI{3}{mm} and \SI{5}{mm} mono-disperse assemblies respectively. Stain localization is seen distributed throughout the assembly. It is difficult to demarcate the shear bands due to the small value of the applied stroke. \cref{fig:seg_sf} shows the same strain distribution seen above this time for the segregated assembly. Here, an interesting pattern emerges. Diffused strain distribution is seen in the region comprising \SI{3}{mm} particles. Comparatively, the regions comprising the \SI{4}{mm}, \SI{4.5}{mm} and \SI{5}{mm} particles are almost devoid of this strain component. This phenomenon is unexpected as both layers are bearing the same load, leading to the expectation that the assembly with lower load-bearing capacity would undergo more shear strain. By that train of thought, at least for the assembly comprising the \SI{3}{mm} and \SI{5}{mm}, we would expect the \SI{5}{mm} layer to undergo more shear strain. This is because, in the mono-disperse case, it has been observed that the \SI{5}{mm} particle assembly was shown to have a lower load-bearing capacity. One more observation, which implies the preceding observation, is that the interface between the two layers of particles seems to act as a barrier for strain localization. While this effect has almost been taken for granted till now, we don't have an explanation for why this happens instead of strain being distributed relatively uniformly throughout the assembly across the interface.


\begin{figure}[H]
     \centering
     \begin{subfigure}{0.5\textwidth}
         \centering
         \caption{Mono sized assemblies}
         \includegraphics[width=0.8\textwidth]{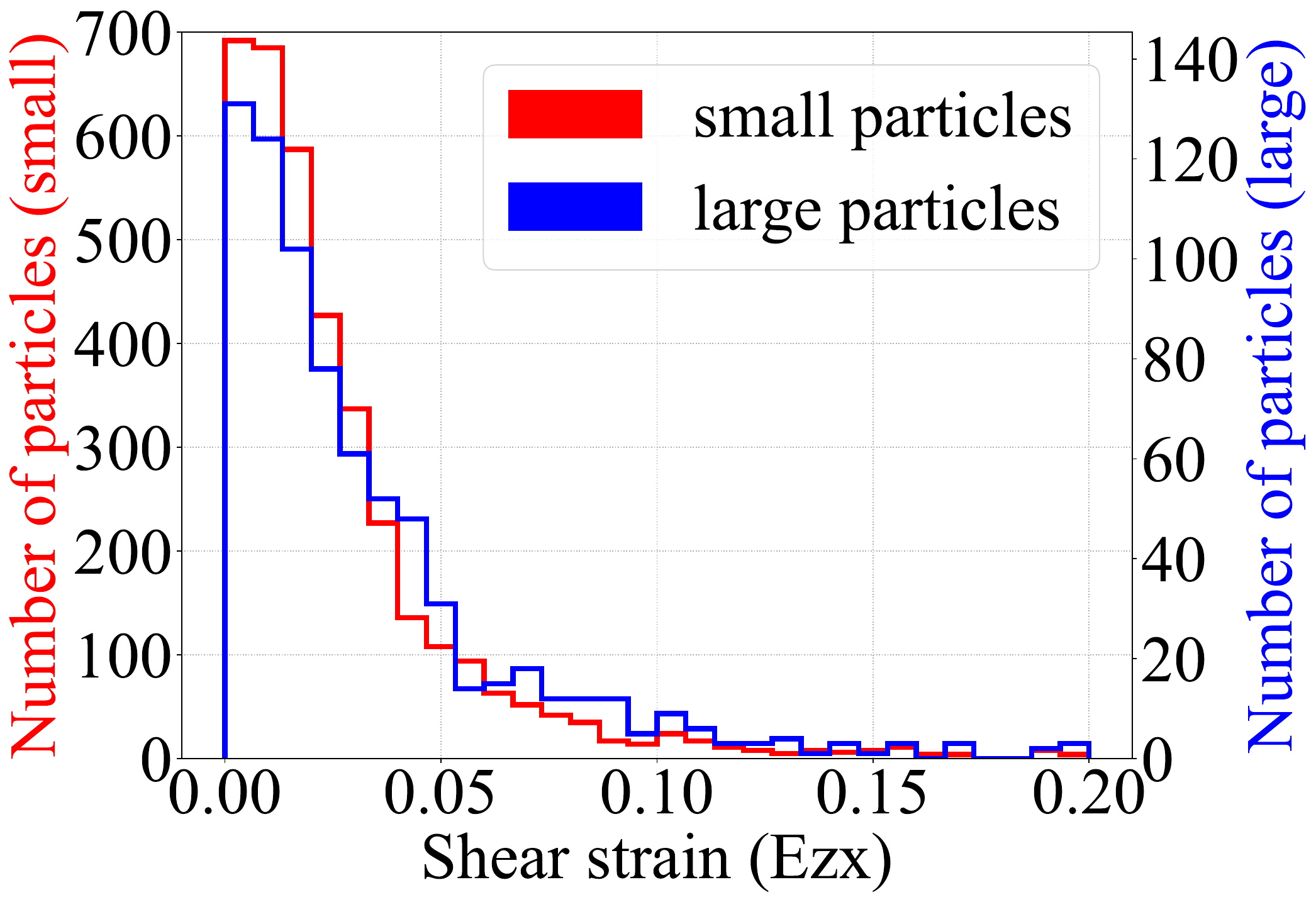}
         
         \label{fig:hist1}
     \end{subfigure}%
     \begin{subfigure}{0.5\textwidth}
         \centering
         \caption{\SI{3}{mm}/\SI{4}{mm} segregated assembly}
         \includegraphics[width=0.8\textwidth]{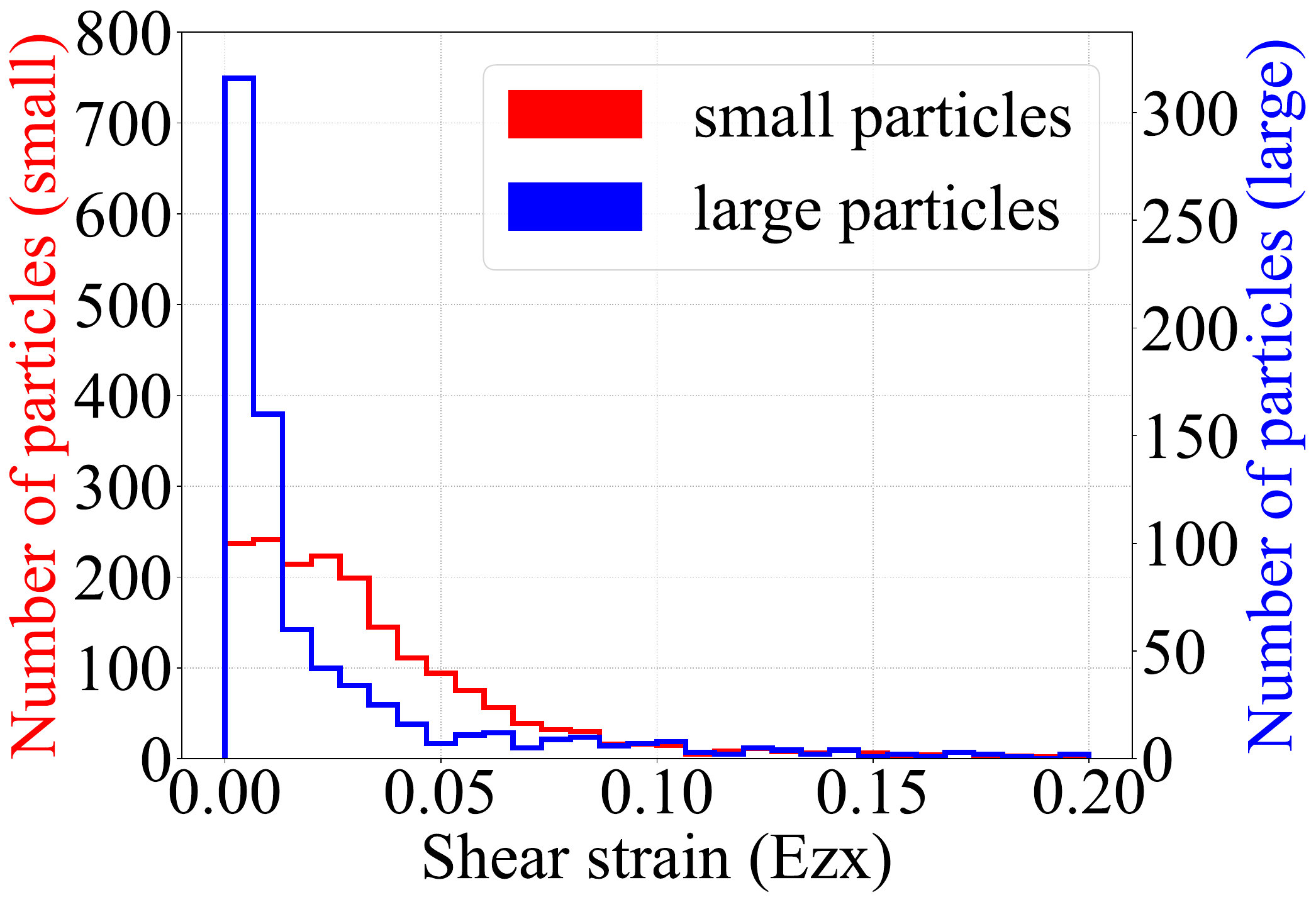}
         
         \label{fig:hist2}
     \end{subfigure}
     
     \begin{subfigure}{0.5\textwidth}
         \hspace{2cm}
         \centering
         \caption{\SI{3}{mm}/\SI{4.5}{mm} segregated assembly}
         \includegraphics[width=0.8\textwidth]{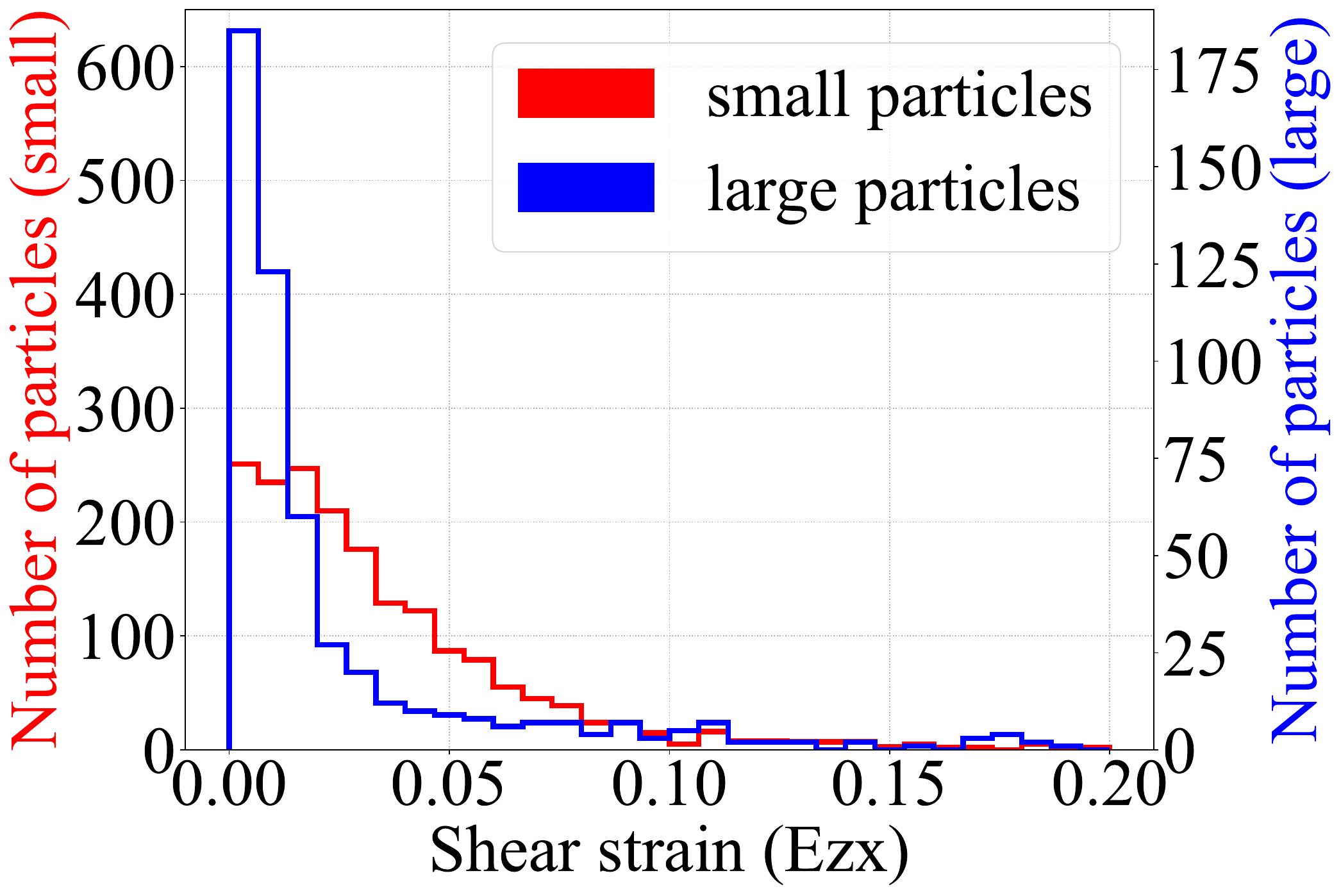}
         
         \label{fig:hist3}
     \end{subfigure}%
     \begin{subfigure}{0.5\textwidth}
         \centering
         \caption{\SI{3}{mm}/\SI{5}{mm} segregated assembly}
         \includegraphics[width=0.8\textwidth]{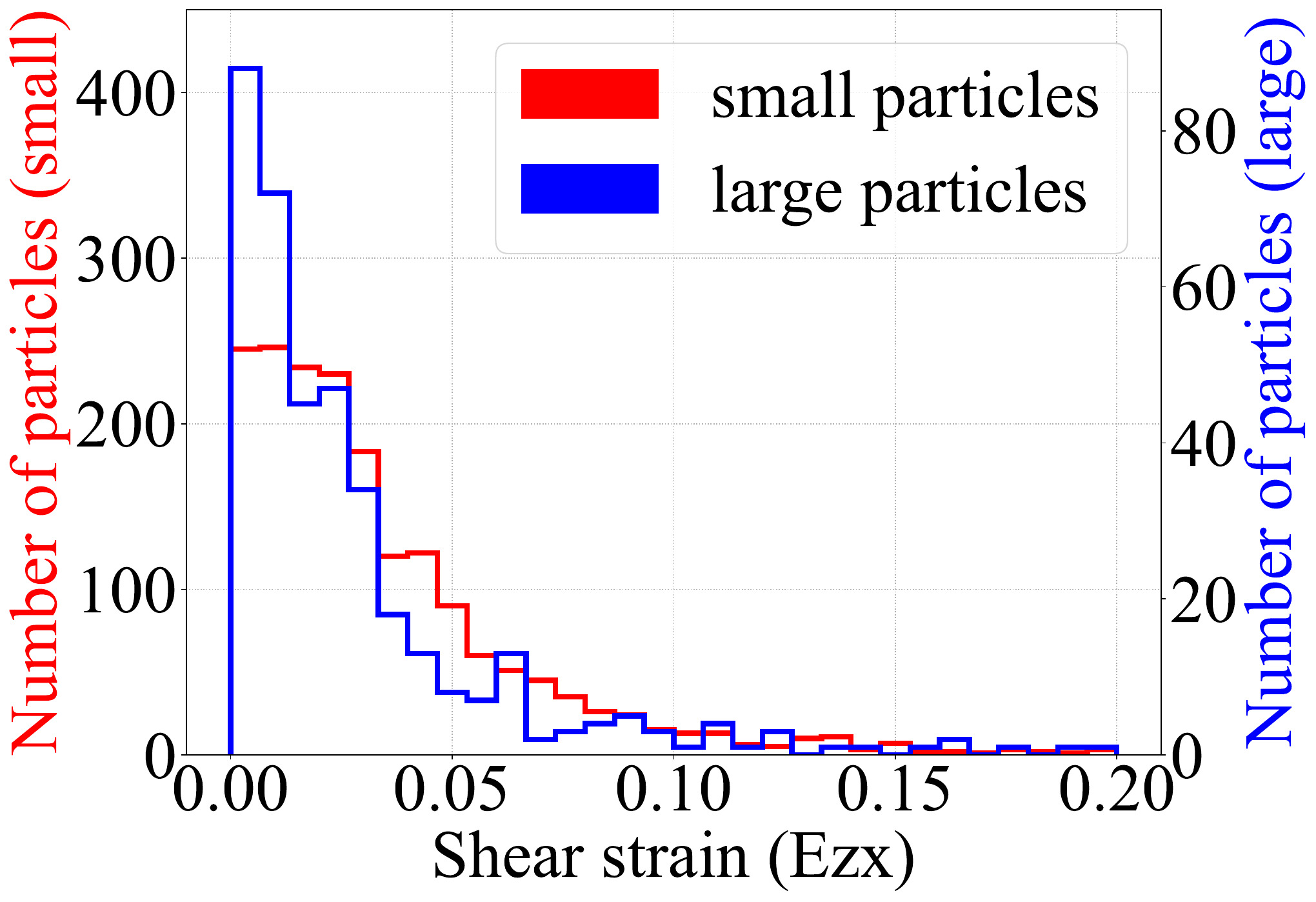}
         
         \label{fig:hist4}
     \end{subfigure}
        \caption{Histograms comparing the distribution of the $zx$ component of Green strain tensor.}
        \label{fig:hists}
\end{figure}

To quantitatively compare this phenomenon, a histogram comparison of the distribution of $zx$ component of the Green strain tensor can be seen in~\cref{fig:hists}. The left y-axis corresponds to the number of small~(\SI{3}{mm}) particles, and the right y-axis corresponds to the large particles. Both the y-axes have been scaled proportional to the total number of particles of the particular kind. This ensures normalization with respect to the number of particles. \cref{fig:hist1} compares the strain amongst the mono-disperse assemblies. We see that between the strain values of 0 and 0.03, the \SI{3}{mm} particles are marginally more in proportion, and the inverse is true post the strain value of 0.3. That being said, the strain distribution is fairly similar in the two cases. However, the same cannot be said for the segregated assemblies. \cref{fig:hist2,fig:hist3,fig:hist4} compares the strain distribution between the \SI{3}{mm} and the \SI{4}{mm}, \SI{4.5}{mm} and \SI{5}{mm} particles in the segregated assembly respectively. In all these cases, a higher number of larger particles undergo strain less than 0.02. Above that value, the \SI{3}{mm} particles occupy a significantly larger proportion. We reiterate that the comparison is not between the actual number of particles but the proportion of particles. This comparison is possible in the graphs because the y-axes have been scaled in proportion to the total number of particles of the particular kind. This re-confirms our previous observation about the \SI{3}{mm} particles preferentially undergoing larger shear strain compared to the larger particles. The average shear strain of the large and small particles for the various assemblies can be seen in ~\cref{tab:my_label}. The difference in the average shear strain between the small and large particles is also shown. For the mono-sized assemblies~(\SI{3}{mm} and \SI{5}{mm}), the larger particles have a marginally higher average shear strain compared to the smaller particles by an amount of 0.6. Surprisingly, for the segregated assembly comprising the \SI{3}{mm} and \SI{5}{mm} also, the average shear strain difference is 0.6, except the \SI{3}{mm} particles have the higher shear strain. Amongst the segregated assemblies, the average shear stain difference seems inversely related to the size of the larger particles.  

\begin{figure}[htb]
     \centering
     \begin{subfigure}{0.24\textwidth}
         \centering
         \includegraphics[width=\textwidth]{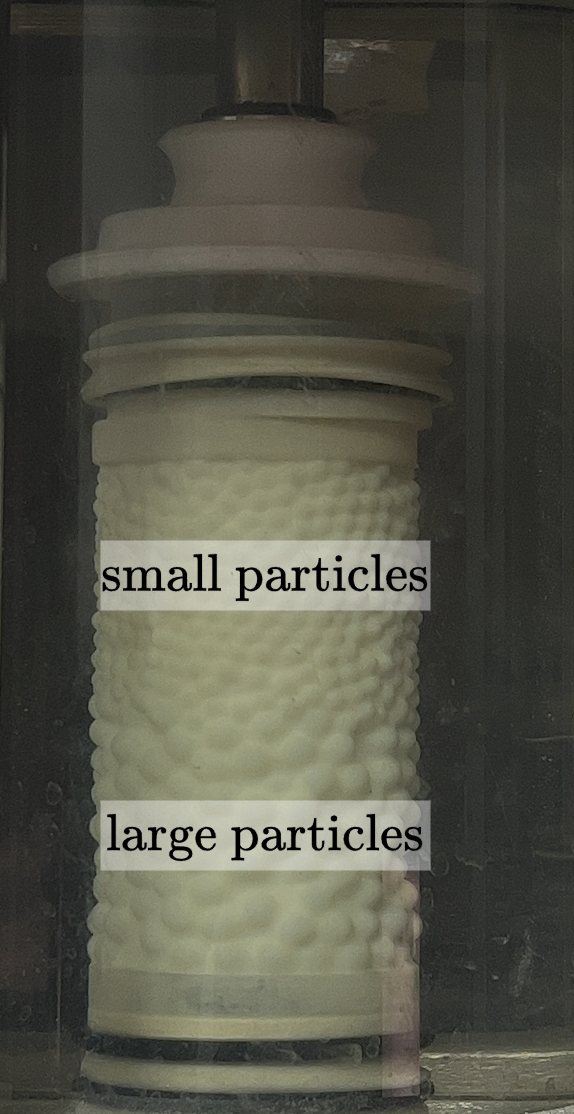}
         \caption{}
         \label{fig:seg1}
     \end{subfigure}
     \begin{subfigure}{0.244\textwidth}
         \centering
         \includegraphics[width=\textwidth]{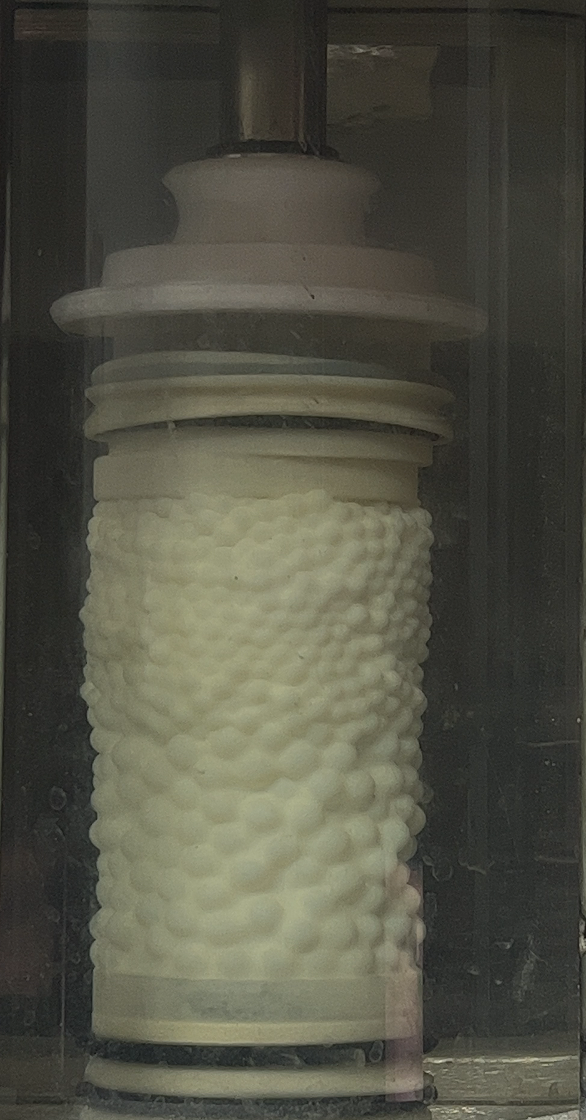}
         \caption{}
         \label{fig:seg2}
     \end{subfigure}
     \begin{subfigure}{0.24\textwidth}
         \centering
         \includegraphics[width=\textwidth]{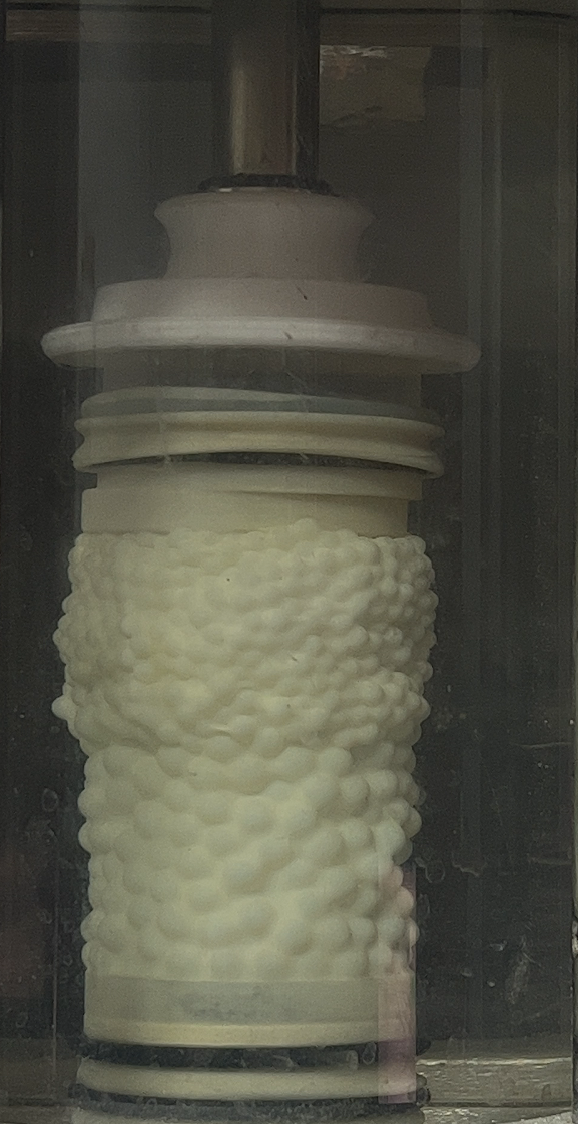}
         \caption{}
         \label{fig:seg3}
     \end{subfigure}
     \begin{subfigure}{0.24\textwidth}
         \centering
         \includegraphics[width=\textwidth]{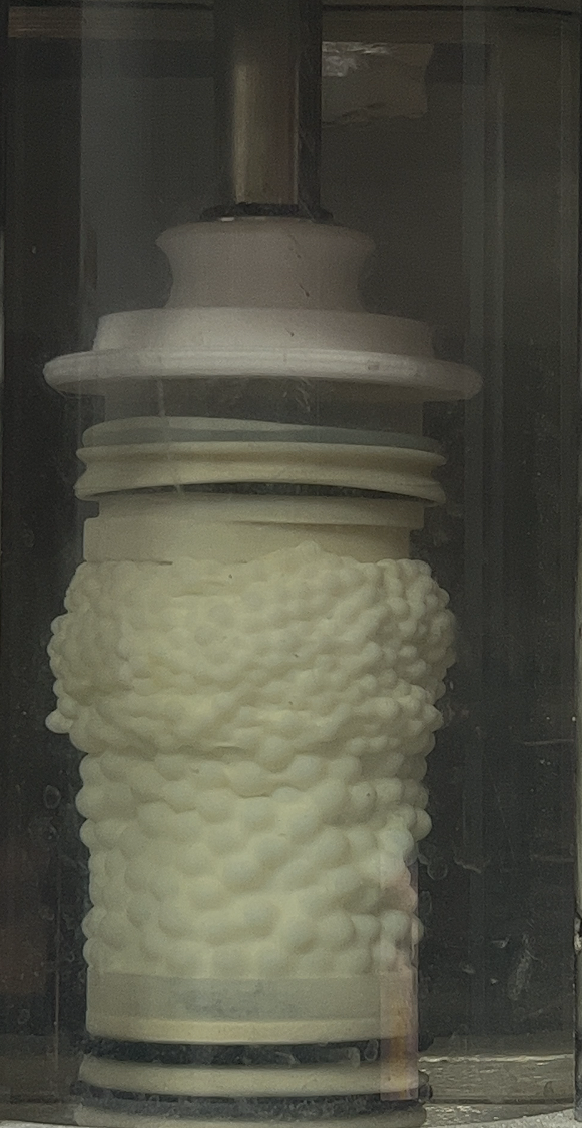}
         \caption{}
         \label{fig:seg4}
     \end{subfigure}
        \caption{Snapshots of the experimental triaxial test of the segregated assembly at~(a) initial state~(b) \SI{5}{mm} stroke~(c) \SI{10}{mm} stroke~(d) \SI{15}{mm} stroke. It can be observed that the smaller particles, occupying the upper half, bulge out, while the larger particles do not experience such bulging.}
        \label{fig:segexcp}
\end{figure}

Some visual observations can be drawn from the experimental triaxial test of the segregated assembly comprising \SI{5}{mm} and \SI{3}{mm}. As the specimen is loaded, we see bulging only in the region containing the \SI{3}{mm} particles as seen in \cref{fig:segexcp}. It is seen that \SI{5}{mm} particles undergo almost no lateral deformation.

\begin{table}
    \centering
    \caption{Comparison of the average value of the $zx$ component of the Green strain tensor~($\Bar{E}_{zx}$) between the large and small particles in all the samples that have been simulated. For the mono-sized samples, the average value corresponds to the average in the respective sample. The errors shown are the standard error values calculated from the standard deviation of the strain distribution.}
    \begin{tabular}{lccc}
    \hline
         Assembly type&$\Bar{E}_{zx}$ small&$\Bar{E}_{zx}$ large& Difference(small-large)\\
    \hline
        Mono-sized(5mm/3mm) & 0.026 $\pm$ 0.0005 & 0.032 $\pm$ 0.0015 & -0.006 $\pm$ 0.0016\\
        Segregated(4mm/3mm) & 0.035 $\pm$ 0.0008 & 0.022 $\pm$ 0.0014 & 0.013 $\pm$ 0.0016\\
        Segregated(4.5mm/3mm) & 0.035 $\pm$ 0.0008 & 0.026 $\pm$ 0.0015 & 0.009 $\pm$ 0.0017\\
        Segregated(5mm/3mm) & 0.033 $\pm$ 0.0007 & 0.027 $\pm$ 0.0015 & 0.006 $\pm$ 0.0017\\
    \hline
    \end{tabular}
    
    \label{tab:my_label}
\end{table}

\clearpage
\section{Conclusion}
\label{sec: conclusion}
The effect of segregation on the strain localization in granular assemblies subjected to triaxial loading has been studied. The discrete element method has been employed to simulate the triaxial loading of granular assemblies. Hexagonally arranged bonded particles were used to model the membrane enveloping the sample. A position-dependent force on the membrane particles was used to apply confinement to the assembly. A mesh reconstruction method has been employed to verify if the correct confinement has been applied to the assembly. Some of the simulations have been verified with experiments by comparing the macroscopic load response, which was found to be in good agreement. The average particle Green strain was calculated using Delaunay triangulation, from which the $zx$ component of the strain was shown along the y-normal plane passing through the centre of the assembly. Finally, the following conclusions have been drawn.
\begin{enumerate}
    \item The position-dependent body-force-based confinement can be used to accurately apply confinement, especially in the initial phase of loading. 
    \item When a segregated sample is subject to triaxial loading, it is observed that only the smaller particles undergo lateral deflection, which is translated into a higher amount of shear strain in the region being occupied by the smaller particles. This effect was also visually observed in the experimentally tested segregated sample.
    \item The difference between the average shear strain experienced by the particles in the segregated assemblies decreases as the size of the larger particles increases, mainly due to the increase in shear strain in the larger particles.
\end{enumerate}
\section*{Conflict of interest statement}
The authors declare that there is no conflict of interest.
\label{sec: Coi}

\section*{Acknowledgements}
\label{sec: acknowledgementes}
The authors are grateful to Altair Engineering India for providing funding, technical support and the software required to run the simulations shown in this article. 

\bibliographystyle{unsrt}
\bibliography{references}

\end{document}